\documentstyle[aps,psfig,floats,graphicx]{revtex}

\begin{document}
\newcommand{\be}{\begin{eqnarray}}
\newcommand{\ee}{\end{eqnarray}}
\newcommand\del{\partial}

\newcommand\barr{|}
\newcommand\cita{\cite}
\newcommand\half{\frac 12}
\newcommand{\el}{\nonumber \hfill \\}
\newcommand{\mat}{\left ( \begin{array}{cc}}
\newcommand{\emat}{\end{array} \right )}
\newcommand{\matf}{\left ( \begin{array}{cccc}}
\newcommand{\ematf}{\end{array} \right )}            
\newcommand{\matt}{\left \begin{array}{ccc}}
\newcommand{\ematt}{\end{array} \right )}
\newcommand{\vect}{\left ( \begin{array}{c}}
\newcommand{\evect}{\end{array} \right )}
\newcommand{\Tr}{{\rm Tr}}
\newcommand{\ZGUE}{{\cal Z}}
\newcommand{\nn}{\nonumber }

\setcounter{page}{0}

\wideabs{
\begin{flushright}
SUNY-NTG-03/28 \\
NORDITA-2003-75 HE 
\end{flushright}

\title{Factorization of Correlation Functions and the Replica Limit of \\
the Toda Lattice Equation}

\author {K. Splittorff$^1$ and J.J.M. Verbaarschot$^2$} 

\address{$^1$ Nordita, Blegdamsvej 17, DK-2100, Copenhagen {\O},
  Denmark \\ $^2$ Department of Physics and Astronomy, SUNY, Stony Brook, New
  York 11794, USA\\email:split@alf.nbi.dk and verbaarschot@tonic.physics.sunysb.edu}

\date   {\today}
\maketitle
\begin  {abstract}

Exact microscopic spectral correlation functions 
are derived by means of the replica limit of the Toda lattice
equation. 
We consider both Hermitian and non-Hermitian theories in the Wigner-Dyson
universality class (class A)
and in the chiral  universality class (class AIII). 
In the Hermitian case we rederive two-point correlation functions for
class A and class AIII as well as several one-point correlation functions
in class AIII. 
In the non-Hermitian case the average spectral density 
of non-Hermitian complex random matrices in the weak non-Hermiticity
limit 
is obtained directly from the replica limit of the
Toda lattice equation. In the case of class A, 
this result describes the spectral density of  a disordered
system in a constant imaginary vector potential (the Hatano-Nelson model)
which is known from earlier work.  
New results are obtained
for the average spectral density in the weak non-Hermiticity limit of
a quenched chiral random matrix model
at nonzero chemical potential. 
These results apply 
to the ergodic or $\epsilon$ domain of the quenched 
QCD partition function at nonzero chemical potential.
Our results have been checked against numerical results
obtained from a large ensemble of random matrices. 
The spectral density obtained is different from the result derived by
Akemann for a closely related model, which is given by
the leading order
asymptotic expansion of our result. In all cases, the replica
limit of the Toda lattice equation explains the factorization of 
spectral one- and two-point functions into a product of a bosonic
(noncompact integral) and a fermionic (compact integral) partition function.
We conclude that the fermionic partition functions, the bosonic
partition functions 
and the supersymmetric partition function are all part of a single integrable
hierarchy. This is the reason that it is possible to obtain the supersymmetric
partition function, and its derivatives, from the replica limit of the Toda 
lattice equation.

\end {abstract}\thispagestyle{empty}
}

\widetext
\onecolumn

\section{Introduction} 
\label{sec:intro}

A replicated system is one in which certain fields appear as
exact copies of each other. 
For a given number, say $n$, of replica fields the system may or
may not have a physical realization, but it is nevertheless useful
to study the system for any integer values of $n$. Suppose, as is the
case in this paper, that we wish to calculate the average resolvent 
of some operator $D$,
\be
G(x) \equiv \left\langle {\rm Tr} \frac 1{x+D} \right\rangle,
\ee
and, as is also the case, that it is simpler to evaluate $\langle
{\det}^n(D+x) \rangle$ for integer values of $n$ than it is to evaluate the
average of the resolvent. 
In that case the replica trick allows us to obtain the resolvent
from the replicated system. The fermionic (bosonic) replica trick consists in
evaluating $\langle {\det}^n(D+x) \rangle$ for all positive (negative) 
integers, $n$, and then taking the limit $n\to0$ as follows
\cite{EA}
\be
G(x) =
\lim_{n\to 0}\partial_x \left\langle \frac {{\det}^n(D+x)}n
\right\rangle =\partial_x \left\langle \log\det(D+x)
\right\rangle . 
\ee 
The tricky part arises because 
the use of the replica trick assumes that an analytic continuation from 
positive or negative integers is uniquely defined.
It has long been known 
\cite{EJ,VZR,critique,KM,lerner,zirn,kim,DV1,ADDV,poulCERN,Levinsen}
that this is the case for perturbative calculations such as for the expansion
of the resolvent in an asymptotic series in $ 1/x$. Problems are known to
occur when one attempts to derive exact 
nonperturbative 
results for the quenched average
resolvent \cite{critique,zirn}. This is one of the reasons that  
alternative methods were introduced 
to evaluate quenched averages. Most notably, the
supersymmetric method \cite{Efetov,VWZ}, 
where the resolvent is obtained from the generating function
\be 
G(x) =
\left .\partial_J \left\langle \frac {{\det}(D+x+J)}{\det(D+x)}
\right \rangle\right |_{J=0},
\ee  
has been very successful. Among other alternatives to the replica trick, 
we only mention the Keldysh method \cite{altland} by 
means of which some nonperturbative results have been derived.
 
The purpose of this
paper is to show that nonperturbative results can be obtained
by means of the replica trick as well.

A determinant raised to an integer power 
inside a statistical average appears in a wide range of theories. 
In disordered condensed matter
systems and random matrix theory, one is usually interested in quenched
disorder where such a determinant enters in the generating function of
Greens functions as discussed in our example of the resolvent.
In Quantum Chromodynamics
Dynamics (QCD), on the other hand, the physical theory is an average over
gauge field configurations of 
a fermionic determinant  raised to the number of flavors.
In lattice gauge simulations, because of computational reasons, 
one often is forced to work with quenched averages (i.e. ignoring
the fermion determinant) which are
not  realized in nature. 
Both in disordered systems and in QCD the low-energy effective theory
of the quenched system is either a supersymmetric nonlinear sigma model 
or a nonlinear sigma model for replicated flavors.  
For perturbative calculations, both
theories are completely equivalent   
\cite{critique,kim,poulCERN,Levinsen,TV24,BG}. 
To prefer one above the other is just
a matter of convenience.  
On the other hand, nonperturbative
results have only been derived by means of the supersymmetric 
method 
\cite{Efetov,VWZ,simons,wegner-super,cartan-zirnbauer,OTV,DOTV}. 
 
 The replica trick has met substantial
critique \cite{critique,zirn} in nonperturbative applications. 
For instance, it was shown that bosonic 
and fermionic replica limits lead to different results for 
the two point function of the Gaussian unitary ensemble and 
that neither limits lead to the correct result \cite{critique}. 
The original critique 
\cite{critique} motivated an insightful and challenging debate 
\cite{KM,lerner,zirn,kim,DV1,ADDV,kanzieper,shinsuke,Yan} the details
of which have been  accounted for in the 
introduction of \cite{SplitVerb}.  
The debate lead to the generally accepted belief that, although the replica 
trick correctly reproduces the asymptotic expansion of the microscopic
spectral correlation functions, it can not be trusted for
 nonperturbative calculations except in 
a few cases, when the series can be 
resummed \cite{shinsuke} or simply terminates \cite{zirn,DV1}.
However, generally this is not the case, and the asymptotic series
leaves much to be desired. For example, since an asymptotic series 
does not determine the cut structure of an analytic function, it is not
possible to obtain the density of eigenvalues from the discontinuity of the
resolvent across its cut.

Over the past few  years intimate connections have been established between 
random matrix theory, integrable systems, the theory of $\tau$-functions,
Painlev\'e equations and unitary matrix integrals
\cite{Okamoto,Perk,thedutch,Kharchev,mmm,Poulpart,kim,dennisvir,FW1,FW2,Witte}.
Kanzieper \cite{kanzieper02} exploited this connection to obtain exact
analytical results by means of the replica limit
of Painlev\'e equations for the generating function of spectral correlation
functions.
 He noted that the replica index only
appears as a coefficient in the associated Painlev\'e equation.
To leading order in the replica index this equation 
is solvable and, with appropriate boundary
conditions, exact analytical results for spectral 
one- and two-point functions functions could be derived.

The ground breaking work of Kanzieper was followed up by the present authors
\cite{SplitVerb} by showing that the replica limit can be taken 
in the Toda lattice equation rather than in the
Painlev\'e equation. This method has two advantages. First, 
it automatically reveals the factorization of correlation functions
or derivatives of correlation functions in terms of a
product of a fermionic and a bosonic partition function. 
Second, in the examples we have considered, either no differential 
equation had to be solved or the desired result followed from the
solution of a simple first order (inhomogeneous) differential equation.
The factorization property, which had not been appreciated before, 
explains precisely why the 
original formulation of the replica trick is bound to be tricky:  
Fermionic partition functions are given by compact integrals which are
analytic functions of the argument for all integer values of the
replica index $n$. It is therefore a mystery how the replica limit,
$n\to0$,  can be a nonanalytic function. This point of the
original critique \cite{zirn} is dealt with automatically in the replica 
limit of the Toda lattice equation which involves both fermionic and
bosonic partition functions. The bosonic partition function is
determined by a noncompact integral and carries the nonanalytic
information.  
 
The Toda lattice is well-known from the theory of exactly solvable
systems \cite{todabook,korepinbook}. It consists of masses interacting
in one dimension through an exponential potential \cite{todabook}.
The Hamilton equations of motion are known as the Toda lattice equation
and its solutions give the position of the masses as a function of
time. The system is integrable and one particular solution can be
written as a unitary matrix integral which, among others, is the
QCD partition function in the $\epsilon$ regime. 
Similar connections exist for all of the classical random matrix
ensembles (see for example \cite{ForresterBook}).
Remarkably, the replica trick also enters in integrable systems 
related to matrix models for
supersymmetric Yang-Mills theories \cite{deboer}, but it is not known
if there is a relation with our approach.

In \cite{SplitVerb} we derived the two-point function of the Gaussian unitary
ensemble and the quenched and partially quenched one-point functions
of the chiral Gaussian unitary ensemble by means of the
replica limit of the Toda lattice equation. 
The purpose of the present paper 
is to show that such strikingly simple results apply more
generally. We will consider three different physical systems: 
the QCD partition function, the Hatano-Nelson model \cite{HN}
and QCD at nonzero chemical potential. The Hatano-Nelson model is a disordered
system in an imaginary vector potential that
enters in the same way as a baryon chemical potential in the Euclidean
QCD partition function. 
In both systems, the symmetries of the generating function are broken
spontaneously. The corresponding Goldstone bosons interact
according to a nonlinear sigma model that is completely determined by
the pattern of spontaneous and explicit symmetry breaking \cite{GL}. In QCD, this
is the well-know chiral Lagrangian which can be extended to include a nonzero
chemical potential \cite{KST}.
For the Hatano-Nelson model a similar effective Lagrangian can be derived
\cite{Efetov}. Its static limit was obtained earlier \cite{FKS} 
as the weak non-Hermiticity limit of  
a complex non-Hermitian random matrix model.

In both theories we will study the nonlinear sigma model in the ergodic
domain which, in QCD, is also known as the $\epsilon$-regime, where the
contribution from the zero-momentum modes (or constant fields) dominate
the partition function. In this domain, the kinetic term of the
$\sigma$ model can be ignored. The corresponding energy scale, which is  known
as the Thouless energy, is the quark mass  for which the Compton wave
length of the Goldstone bosons is of the order of the size of the box.
In this regime, all theories with  a given symmetry breaking pattern
are universally described by the same partition function. Perhaps the simplest
theories in this universality class are  
random matrix theories. In random matrix theories one seeks to
describe average spectral properties on the scale of the average eigenvalue
spacing in the limit of large matrices. It has been shown
that spectral correlation functions in this microscopic limit are
universal in the sense that they only depend on the symmetries of the
random matrix model and not on its detailed form (see for example 
\cite{hacken,universality}). We now
understand that \cite{OTV,cartan-zirnbauer} this universality is a
direct consequence of the uniqueness of the static part of the effective
Lagrangian. We will exploit the correspondence between random matrix theory
and the static part of the effective Lagrangian to derive the bosonic
partition partition function for QCD at nonzero chemical potential and
for the Hatano-Nelson model. In both cases, the integration contour
of the bosonic fields in the $\sigma$-model follows naturally from 
the Ingham-Siegal integral of the second kind \cite{Yan}.

The two-point functions of the unitary ensemble, the chiral unitary ensemble
and the spectral density of the Hatano-Nelson model \cite{FKS,Efetov} and
the chiral unitary ensemble at nonzero chemical potential
all share one remarkable property. They 
factorize into a product of a fermionic
and a bosonic partition function. We will show that this is a natural 
consequence of the Toda lattice equation. Indeed,
the corresponding generating functions can be written as a 
determinant of derivatives. Therefore, 
they are  $\tau$-functions of a Toda lattice
hierarchy and satisfy a Toda lattice equation.
Our result for the chiral unitary ensemble at nonzero chemical potential
is a genuinely new result which is in complete
agreement with results obtained from a numerical matrix
diagonalization. The leading term of an asymptotic expansion of our result 
and the result for a closely related model for
non-Hermitian spectra with a chiral symmetry \cite{gernotSpectra} turn
out to be the same.

\vspace{4mm}

The organization of this paper is as follows. As a warm up we start
in sections \ref{twotwo} and \ref{sec:PQres}
with simple examples that have  already been discussed
in \cite{SplitVerb}. The two-point 
function of the chiral unitary ensemble is discussed in section
\ref{sec:2pf}. 
Turning to the complex spectra
of non-Hermitian operators we consider the case of a disordered system
in a nonzero imaginary vector
potential in section \ref{sec:EfetovCase} and QCD at non zero chemical
 potential in section \ref{sec:phqchGUE}. Concluding remarks are made
in section \ref{sec:conc}. Our notation as well as additional 
technical details are explained in several appendices
(A through D).


\section{Toda Lattice Equation for Hermitian Theories} 

In this section we start with three simple examples of the replica limit
of the Toda lattice equation, i.e. 
the two-point function of the Unitary Ensemble (UE), the 
resolvent of the chiral Unitary Ensemble (chUE)  
in the quenched case for
topological charge $\nu$, and in the case of $N_f$ massless flavors
in the sector of zero topological charge. 
The first two cases were already discussed in
\cite{SplitVerb}. In the second part of this section we discuss
the calculation of the two-point function of the chUE. In this case the  
Toda lattice equation is obtained from a new representation of the finite
volume partition function which might also be  useful in other applications.

\subsection{The Two-Point Function of the Unitary Ensemble}
\label{twotwo}
This section discusses the simplest example of the replica limit
of the Toda lattice equation \cite{SplitVerb}, namely the
connected two-point function of the Gaussian Unitary Ensemble (GUE).
Its generating function for $n$ fermionic replicas
is given by
\be 
\int dH {\det}^n(H +E_1) {\det}^n(H +E_2)
e^{-\frac{N}{2}{\rm Tr} H^\dagger H}.
\ee
Following the usual reduction, 
this partition function can be written as a nonlinear $\sigma$-model. 
In the large $N$-limit, at fixed $r=N(E_2-E_1)$ near 
the center of the spectrum, the eigenvalue representation of
this  $\sigma$-model further simplifies to  
(see eq. (1.2) of \cite{critique}) 
\be 
 \ZGUE_n(ir) =  n!\int_{-1}^{1}\prod_{k=1}^n d\lambda_k 
e^{-ir\lambda_k} \Delta^2(\lambda).
\label{ZGUE}
\ee  
Here, $\Delta(\lambda)$ is the Vandermonde determinant
$\Delta(\lambda) = \prod_{k<l} (\lambda_k -\lambda_l)$.
The connected 
two-point correlation function is then obtained simply as the 
replica limit
\be
G(r) = -\lim_{n\to 0} \frac 1{n^2} \partial_r^2 \
\log \ZGUE_n(ir).
\ee
 The generating
function (\ref{ZGUE}) can be rewritten as
\be 
 \ZGUE_n(r) = (n!)^2\det[\partial_r^{i+j} \ZGUE_1(r)]_{i,j=0,\ldots,n-1}.
\ee

In a similar way the  large $N$ limit of the bosonic partition can be
reduced to  
\be
 \ZGUE_{-n}(ir) &=& \frac 1{n![(n-1)!]^2} 
\int_{1}^{\infty}\prod_{k=1}^n d\lambda_k 
e^{ir\lambda_k} \Delta^2(\lambda).
\ee
where the sign of the exponent is determined by the sign of the imaginary
part of $r$. The $r$ dependence can be scaled outside the integral resulting
in
\be
 \ZGUE_{-n}(r)       
&=&  \frac{e^{nr}}{r^{n^2}} \prod_{k=1}^{n-2}(k!)^2.
\ee
This expression can be rewritten as
\be
 \ZGUE_{-n}(r) = \frac 1{[(n-1)!]^2}\det[\partial_r^{i+j} 
\ZGUE_{-1}(r)]_{i,j=0,\ldots,n-1}.
\ee
Therefore, both the fermionic and the bosonic partition function
satisfy
the Toda lattice equation \cite{kanzieper,SplitVerb}
\be
-\ZGUE_n^2(ir)\partial_r^2 \log \ZGUE_n(ir) = \frac{n^2}{(n+1)^2} 
\ZGUE_{n+1}(ir) \ZGUE_{n-1}(ir).
\ee
In the replica limit, the l.h.s. of this equation is $n^2$ times 
the connected two-point function whereas the r.h.s. correctly gives 
\be
G(r) = \ZGUE_1(ir) \ZGUE_{-1}(ir)
= \frac{2\sin r}{ r} \frac {ie^{ir}}{r}.
\ee
We thus find that the factorization of the two-point correlation 
function into a bosonic and a fermionic partition function is not
accidental but rather a consequence of 
the relation between random matrix theories and integrable hierarchies.
The two-point function of the GUE can also be obtained from a solution
of the Painlev\'e V equation \cite{kanzieper}.
Another feature of this result emphasized in \cite{zirn} is that
the asymptotic series in $1/r$ terminates. This explains \cite{zirn}
that in this case the exact result could be obtained 
\cite{KM} by including a subleading saddle point manifold
in the standard replica approach.

\subsection{The Partially Quenched Resolvent in QCD}
\label{sec:PQres}

In this section we discuss the replica limit of the 
Toda lattice equation for the generating function of the
partially quenched resolvent 
in chUE. 
As a simple extension of the 
results presented in \cite{SplitVerb} we obtain the 
microscopic limit of the partially quenched 
resolvent for an arbitrary number of flavors. 

The partition function of the chiral Gaussian Unitary Ensemble (chGUE) 
with $n$ fermionic replica flavors
with mass $m$ and $N_f$ additional fermionic flavors with masses $m_1,
\cdots, m_{N_f}$ is defined by 
\be
Z^{(\nu)}_{n,N_f}(m,\{m_i\}) = 
 \langle {\det}^n (D +m) \prod_{k=1}^{N_f}\det(D+m_k) \rangle .
\label{ZchGUE}
\ee
The Dirac matrix is given by
\be 
D\equiv\left(\begin{array}{cc} 0 & iW \\ iW^\dagger & 0
\end{array}\right) ,
\label{D}
\ee
where $W$ is a rectangular $l\times(l+\nu)$ matrix so that $D$ has
exactly $\nu$ zero eigenvalues. The average is over the probability
distribution of the matrix elements of $W$ with the weight 
\be
P(W)=e^{-\frac 12 N {\rm Tr}(WW^\dagger)}. 
\label{weightW}
\ee
with $N=2l+\nu$. 
We will consider the microscopic limit where the thermodynamic limit
is taken for fixed $x\equiv m N$ and $x_k \equiv m_k N$. 
In this case the partition function reduces to the unitary matrix integral
\cite{SV} 
\be
Z_{n,N_f}^{(\nu)}(x,\{x_i\}) = \frac 1{(2\pi)^{n(n+1)/2}}
\int_{U \in U(n+N_f) } \hspace{-12mm} dU {\det}^\nu U e^{\frac 12
{\rm  Tr}[M^\dagger U +M U^\dagger]},
\label{pq}
\ee
where $M={\rm diag}(x,\cdots,x,x_1,\cdots,x_{N_f})$ is the mass matrix.
The normalization factor has been included for later convenience. 
The partially quenched resolvent, which we will derive below, is given by
\be
G(x,\{x_i\}) = 
\lim_{n \to 0} \frac 1n \partial_x \log Z_{n,N_f}^{(\nu)}(x,\{x_i\}).
\ee 
 The unitary matrix integral
(\ref{pq}) is the partition function of QCD \cite{SV,GLeps,LS} in the
$\epsilon$-regime where it coincides with the microscopic limit
of the chGUE partition function (\ref{ZchGUE}). 
Partition functions for different number of
flavors satisfy a recursive relation know as the Toda lattice
equation \cite{Kharchev}  
\be
\label{genToda}
\sum_{k=0}^{N_f} 
\left . x_0\partial_{x_0}x_k\partial_{x_k} 
\log Z^{(\nu)}_{n,N_f}(x_0,\{x_k\}) \right|_{x_0 =x} 
& = & 
 n x^2\frac{ Z^{(\nu)}_{n+1,N_f}(x,\{{x_k}\})
Z^{(\nu)}_{n-1,N_f}(x, \{x_k\})}
{[Z^{(\nu)}_{n,N_f}(x,\{x_k\})]^2} .
\ee
For degenerate masses $x_k$ the sum over the derivatives is only over
different masses.

The bosonic chGUE partition function is defined as in (\ref{ZchGUE}) but
with negative integer values of $n$.  
Its axial symmetry is not $U(n+N_f)$ but
rather $Gl(n+N_f)/U(n+N_f)$ \cite{cartan-zirnbauer,OTV,DV1}. 
Therefore, the Goldstone manifold for
the bosonic partially quenched partition function is
 $Gl(n+N_f)/U(n+N_f)$ rather than  $U(n+N_f)$, i.e. the coset of positive
definite matrices. In this way the integration manifold remains Riemannian
\cite{cartan-zirnbauer} so that all integrals are  convergent for
positive masses. 
As bosonic partition function we thus obtain \cite{OTV,DV1}
\be
Z_{-n}^{(\nu)}(x,\{x_i\}) = \frac 1{C_{-n}}
\int_{Q \in Gl(n)/U(n) } \hspace{-12mm} dQ{\det}^\nu Q e^{-\frac 12
{\rm  Tr}[M^\dagger Q +M Q^{-1}]} .
\label{pq2}
\ee 
If we parameterize $Q$ as
\be
Q = U {\rm diag}(e^{s_k}) U ^\dagger,\qquad U^\dagger U = 1
\ee
the invariant measure is given by
\be
dQ = \prod_{k<l} (e^{s_k} - e^{s_l})(e^{-s_k} - e^{-s_l})
\prod_k ds_k dU .
\ee
Using the Ingham-Siegel integral \cite{Yan} this result can be derived 
directly from the bosonic random matrix partition function \cite{FSV}.
For zero topological charge this is discussed in \cite{Yan} and  
the simplest nontrivial example is given in \ref{App:Zm1mu=0}. The
explicit form of the partition function with any number of fermions 
and bosons in the topological sector $\nu$ conjectured in
\cite{SplitVerb} was proven rigorously in \cite{FA} based on a general
result in \cite{FSgen}. 

In \cite{SplitVerb}, the $n\to0$ limit of the Toda lattice equation
(\ref{genToda}) was solved for $N_f =0$ and $N_f =1$. Before we 
consider the case with
$N_f$ massless flavors and zero topological charge, 
we first review 
the quenched limit ($N_f =0$) of (\ref{genToda}) in the
sector of topological charge $\nu$.  
In this case, (\ref{genToda}) simplifies to 
\be
[x \partial_x]^2 \log Z_n^{(\nu)}(x) = n x^2 
\frac{Z^{(\nu)}_{n+1}(x) Z^{(\nu)}_{n-1}(x)}{[Z^{(\nu)}_n(x)]^2}.
\label{todanu}
\ee
Collecting terms of ${\cal O}(n)$ in the replica limit of this Toda
lattice equation we find 
\be
\partial_x [x G(x)] = x Z^{(\nu)}_1(x) Z^{(\nu)}_{-1}(x).
\label{inhomogeneous}
\ee
Again, we observe a factorization property of the resolvent.
For topological charge $\nu$ we have that 
\footnote{
The normalization factor of the bosonic partition function is chosen
such that the large $x$ limit of both sides of (\ref{inhomogeneous}) 
is the same.
For large $x$ we have that $G(x) = 1$  whereas the large $x$ behavior
of the r.h.s. of (\ref{inhomogeneous}) is obtained from the leading order
asymptotic expansion of the Bessel functions.}
\be
Z^{(\nu)}_1    (x) =  I_\nu(x) \quad {\rm and } \quad 
Z^{(\nu)}_{-1} (x) = 2K_\nu(x).
\ee
The general solution of the inhomogeneous differential equation
(\ref{inhomogeneous}) is given
by
\be
x G(x) = a + x^2 (K_\nu(x)I_\nu(x) + K_{\nu-1}(x) I_{\nu+1} (x)),
\ee
where the first term is a solution of the homogeneous differential equation.
The integration constant is fixed by the small $x$ expansion of the 
resolvent which in the sector of topological charge $\nu$ is given by
\be
G(x) \sim \frac \nu x,
\ee
so that $a=\nu$. 
This ends the derivation of the quenched resolvent in the topological
sector $\nu$. The result is in exact agreement with the known result
\cite{Vplb}.

As a final introductory example, let us now consider the case of $N_f$
massless flavors and zero topological charge. 
The general Toda lattice equation (\ref{genToda}) reduces to 
\be\label{todanf0}
\left. \lim_{n\to 0} \frac 1n (x\partial_x + y\partial_y) x\partial_x 
\log Z^{(0)}_{n,N_f}(x,y)\right |_{y=0}
=
 x^2 \frac {Z_{1,N_f}^{(0)}(x,0)Z^{(0)}_{-1,N_f}(x,0)}
{[Z^{(0)}_{0,N_f}(x,0)]^2} .
\ee
The partition function $Z_{1,N_f}^{(0)}(x,0)$ can be obtained by using
the flavor-topology duality relation \cite{cam97} and is given by
\be 
Z_{1,N_f}^{(0)}(x,0) = x^{-N_f} Z_{1,0}^{(\nu=N_f)}(x) =
x^{-N_f} I_{N_f}(x).
\ee
The partition function $Z^{(0)}_{-1,N_f}(x,0)$ has $N_f$ massless
fermionic flavors and one bosonic flavor with mass $x$. Using a
bosonic version of the flavor-topology duality relation we find 
\be
Z^{(0)}_{-1,N_f}(x,0) = x^{N_f} Z^{(\nu=N_f)}_{-1,0}(x) = 2x^{N_f} K_{N_f}(x),
\ee
where the normalization factors are consistent with the
asymptotic behavior of (\ref{todanf0}). 
The $y \partial_y$-derivative of $\log Z^{(0)}_{n,N_f}(x,y)$ vanishes
for $y =0$ so that (\ref{genToda}) simplifies to
\be
\partial_x[x G(x)] = 2 x K_{N_f}(x) I_{N_f}(x).
\ee
As in the quenched case for topological charge $\nu$, the solution is
given by
\be
xG(x) = a +  x^2 [ K_{N_f}(x) I_{N_f}(x) + K_{N_f-1}(x) I_{N_f+1}(x)],\nonumber
\ee
Because we are in the sector of zero topological charge, the resolvent approaches a constant for small $x$ so that the integration constant
$a = 0$. This result agrees with the result obtained by integration
the microscopic spectral density \cite{Vplb}.

\subsection{The Two-Point Function in QCD}
\label{sec:2pf}

In this section we consider the microscopic limit of the chUE partition
function for $m$ flavors with mass $x$ and $n$ flavors with mass $y$.
In the sector of topological charge $\nu$ we will denote this partition
function by $Z^{(\nu)}_{m,n}(x,y)$.
The microscopic
spectral two-point function will be derived 
from the replica limit of the Toda lattice
equation corresponding to this partition function. 
In the first subsection
we will show that the partition function can be written as a 
determinant of derivatives.
In the second subsection we shall
use this   
form to show that $Z^{(\nu)}_{m,n}(x,y)$ satisfies a  Toda lattice
equation and therefore is a $\tau$-function of an integrable 
hierarchy. The replica limit of the Toda lattice equation then
automatically gives us the exact analytical result for 
the two-point function that was obtained earlier \cite{TV-two}
by means of the supersymmetric method.

\vspace{2mm}
\subsubsection{The Generating Function for the Two-Point Function
is a $\tau-$Function}
 
The finite volume partition function for $m$ fermionic flavors with
mass $x$ and $n$ (with $n\geq m$) fermionic flavors  with mass $y$ is given
by \cite{SV,GLeps,LS}
\be 
Z^{(\nu)}_{m,n}(x,y) \equiv \frac 1{T_{m,n}}\int_{ U(n+m)} dU  {\det}^\nu(U) 
\exp[\frac 12 {\rm Tr} \mat x{\bf 1}_m & \\ & y{\bf 1}_n \emat U
   + \frac 12{\rm Tr} \mat x{\bf 1}_m & \\ & y{\bf 1}_n \emat U^\dagger] .
\ee
This partition function can be derived from a random matrix theory
with the symmetries of the QCD partition function \cite{SV} and is also
known as the QCD partition function in the $\epsilon$ regime. 
This integral and its generalizations have been well studied in the literature 
\cite{KSS,Brower,jsv,Tilo,baba}, but the known analytical expressions do not
directly show that 
$Z^{(\nu)}_{m,n}(x,y)$ is a $\tau$-function. In order to write it in
such a form we evaluate the integral over $U$ by decomposing $U(n+m)$
as $U(m)\times U(n)\times [ U(n+m)/(U(n) \times U(m))]$, and
evaluating the integral over the coset first. The normalization factor
is chosen to be
\be
T_{m,n} = (2\pi)^{(n+m)(n+m+1)/2}.
\ee 
Our calculations simplify if we use the parameterization
\be
U = \mat u_1 & \\ & u_2 \emat \mat v_1 & \\ & v_2 \emat  
\Lambda \mat v_1^\dagger & \\ & v_2^\dagger \emat, \label{paramSec}
\ee
where $\Lambda$ is the block diagonal $(m+n)\times(m+n)$ matrix given by
\be
\Lambda &=& \left(\begin{array}{ccc} 
\sqrt{1-\mu^2}& \mu & 0 \\ 
\mu & -\sqrt{1-\mu^2} & 0 \\ 
0 & 0 & -{\bf 1}
\end{array}\right).
\ee
The diagonal matrix $\mu={\rm diag}(\mu_1,\cdots, \mu_m)$, and we will
use the notation that $\lambda_k = \sqrt{1-\mu_k^2}$ with
$\lambda_k\in[0,1]$. The unit matrix ${\bf 1}$ in the lower right hand
corner is of size $n-m$. Furthermore,
$u_1$ and $v_1$ are unitary $m\times m$ matrices, $u_2 \in U(n)$ and
$v_2 \in U(n) / (U^m(1) \times U(n-m))$.  One easily verifies that the total
number of parameters on both sides of (\ref{paramSec}) is the same. The 
Jacobian of this transformation is derived in
\ref{App:Jacobian} and is given by
\be
J = \prod_{1\le k<l\le m} (\lambda_k^2 -\lambda_l^2)^2 
\prod_{k=1}^m (2\lambda_k) \mu_k^{2(n-m)}.\label{jactotText}
\ee
The finite volume partition function can thus be rewritten as (the $v_k$
variables have been eliminated from the exponent by the transformation
$u_k \to v_k u_k v_k^{-1}$)
\be 
Z^{(\nu)}_{m,n}(x,y) = &&\frac 1{m!T_{m,n}}\int_{U(m)}  dv_1
\int_{U(n)/[U^{m}(1) \times U(n-m)]} dv_2 
\int_{U(m)} du_1 \int_{U(n)} du_2 
\int_0^1 \prod_{k=1}^m d\lambda_k          
\nn \\ &&\times 
J(\{\lambda_k\})
{\det}^\nu(u_1 u_2) 
\exp[\frac 12 x {\rm Tr}\Lambda_{11} (u_1 +u_1^\dagger)
+\frac 12 y {\rm Tr}\Lambda_{22} (u_2 +u_2^\dagger)]. 
\label{zr1}
\ee
Here we have introduced the notation
\be
\Lambda_{11}\equiv{\rm diag}(\lambda_1,\ldots,\lambda_m),
\ee
and
\be
\Lambda_{22}\equiv{\rm diag}(-\lambda_1,\ldots,-\lambda_m,-1,\ldots,-1).
\ee

The integrals over $v_1$ and $v_2$ give an overall constant equal to the
volume of the integration manifold. The integrals over $u_1$ and $u_2$ are
well known. They are given by \cite{Brower,jsv,Tilo,baba}
\be
\frac 1{{\rm vol}(U(m))}
\int_{U(m)} du_1  {\det}^\nu(u_1) 
\exp[{\rm Tr} (\frac 12 x\Lambda_{11}u_1 + \frac 12x\Lambda_{11}u_1^\dagger)] 
= C_m\frac {\det [\phi_l(x \lambda_k)]_{k,l = 0, \cdots, m-1}}
           {\Delta(\{(x\lambda_k)^2\})},
\label{Zfinite}
\ee
where
\be
\phi_l(x_k) \equiv (x_k \partial_{x_k})^l I_\nu(x_k),
\ee
and
\be
C_m \equiv 2^{m(m-1)/2} \prod_{k=1}^{m-1} k!\,.
\ee 
The volume of the unitary group is given by (see eg. \cite{Hua})
\be
{\rm vol}(U(n))\equiv \int_{U(n)} dU =
\frac{(2\pi)^{n(n+1)/2}}{\prod_{k=1}^{n-1} k!}.
\ee
A similar formula can be derived for the $u_2$-integral. The only complication
is that $n-m$ diagonal matrix elements of $y\Lambda_{22}$ are degenerate.
To apply the formula (\ref{Zfinite}) we add $\epsilon_k$ to
the degenerate diagonal matrix elements. The
$\epsilon_k \to 0$ limit can then be obtained conveniently by writing the 
Taylor expansion of  
$\phi_l(-y+\epsilon_k)$
to order $n-m$ in $\epsilon_k$ as a product of a matrix containing the
powers of $\epsilon_k$ and a derivative matrix.
This results in
\be 
&&\frac 1{{\rm vol}(U(n))}
\int_{U(n)} du_2  {\det}^\nu(u_2) \exp[{\rm Tr} (\frac 12y\Lambda_{22} u_2
  +\frac 12y\Lambda_{22} u^\dagger_2)]  
\nn \\
&&= 
\frac{C_n}{C_{n-m}}\frac {y^{-(n-m)(n-m-1)/2}} 
{\prod_{k=1}^m(\lambda_k^2y^2 -y^2)^{(n-m)} 
\prod_{1\le k<l\le m} (\lambda_k^2y^2 -\lambda_l^2y^2)}
\det \left | \begin{array}{ccc}
\phi_0(-y\lambda_1) & \cdots &\phi_{n-1}(-y\lambda_1) \\
  \vdots    &         & \vdots \\
\phi_0(-y\lambda_m) & \cdots &\phi_{n-1}(-y\lambda_m) \\
\phi_0(-y) & \cdots &\phi_{n-1}(-y) \\
 \vdots    &         & \vdots \\
\phi_0^{(n-m-1)}(-y) & \cdots &\phi_{{n-1}}^{(n-m-1)}(-y) \end{array} \right |, 
\label{limit}
\ee  
where we have  used the notation
$\phi_{k}^{(l)}(y)\equiv\del_y^l\phi_{k}(y)$. 
All factors in the Jacobian (\ref{jactotText}) except $\prod (2\lambda_k)$
are canceled by the corresponding factors from the $u_1$ and $u_2$ 
integrations. Multiplying the two determinants the partition function
can be written as
\be
Z^{(\nu)}_{m,n}(x,y) =
\frac{ 2^{(m+n)(m+n-1)/2} y^{(n-m)(n-m-1)/2}}
{C_{n} C_m x^{m(m-1)}y^{n(n-1)}}
\det \left | \begin{array}{ccc}    
\int_0^1 d\lambda \,\lambda \phi_0(\lambda x) \phi_{0}(-\lambda y) & \cdots &
\int_0^1 d\lambda \,\lambda \phi_0(\lambda x) \phi_{n-1}(-\lambda y)\\
 \vdots    &         & \vdots \\
\int_0^1 d\lambda \,\lambda \phi_{m-1}(\lambda x) \phi_0(-\lambda y) & \cdots &
\int_0^1 d\lambda \,\lambda \phi_{m-1}(\lambda x) \phi_{n-1}(-\lambda y)\\
\phi_0(-y) & \cdots &\phi_{n-1}(-y) \\ 
 \vdots    &         & \vdots \\
\phi_0^{(n-m-1)}(-y) & \cdots &\phi_{n-1}^{(n-m-1)}(-y) \end{array} \right | .
\ee
By the addition of rows, the derivatives in the last $m$ rows can 
be rewritten in terms of the derivative operators
\be 
\delta_x \equiv x\frac {d}{dx} \,\,.
\ee
This results in the final form for the partition function of the chUE
\be
Z^{(\nu)}_{m,n}(x,y) =
\frac{ 2^{(m+n)(m+n-1)/2}}
{C_{n} C_m x^{m(m-1)}y^{n(n-1)}}
\det \left | \begin{array}{ccc}
\int_0^1 d\lambda \,\lambda I_\nu(\lambda x) I_\nu(-\lambda y) & \cdots &
\delta_y^{n-1} \int_0^1 d\lambda \,\lambda I_\nu(\lambda x) I_\nu(-\lambda y)\\
 \vdots    &         & \vdots \\
 \delta_x^{m-1}\int_0^1 d\lambda \,\lambda I_\nu (\lambda x) I_\nu(-\lambda y) 
& \cdots &  \delta_x^{m-1}\delta_y^{n-1} 
\int_0^1 d\lambda \,\lambda I_\nu(\lambda x) I_\nu(-\lambda y)\\
I_\nu(-y) & \cdots &\delta_y^{n-1} I_\nu(-y) \\ 
 \vdots    &         & \vdots \\
\delta_y^{n-m-1}  I_\nu(-y) & \cdots & \delta_y^{2n-m-2}I_\nu(-y) 
\end{array} \right | .
\label{partf}
\ee
If we denote the $n\times n$ matrix that enters in this partition
function by $A^{(m,n)}$ it can be rewritten as 
\be
 Z^{(\nu)}_{m,n}(x,y) = 
\frac{ 2^{(m+n)(m+n-1)/2}}
{C_{n} C_m x^{m(m-1)}y^{n(n-1)}}
 \det A^{(m,n)}(x,y). 
\label{ZchGUEasTau}
\ee
Of course, $A^{(m,n)}$ is a function of $\nu$ even though we have not 
explicitly indicated it. Before we proceed to show that this partition
function satisfies a Toda lattice equation, let us note that: 
$Z_{1,1}^{(\nu)}(x,y)=2\int d\lambda \,\lambda I_\nu(\lambda x) 
I_\nu(-\lambda y)$. Hence, for $n=m$ we obtain the compact expression
\be
Z^{(\nu)}_{n,n}(x,y)= \frac {2^{n(2n-1)}}{ C_n^2 (xy)^{n(n-1)}}
\det[\delta_x^{i}\delta_y^{j}Z_{1,1}^{(\nu)}(x,y)]_{i,j=0,\ldots,n-1}. 
\ee
Another special case is  $m=0$, where the partition function simplifies to
\be
Z^{(\nu)}_{m=0,n}(x,y) =
\frac{2^{n(n-1)/2}}{C_n y^{n(n-1)}} 
\det \left | \begin{array}{ccc}
I_\nu(-y) & \cdots &\delta_y^{n-1} I_\nu(-y) \\ 
 \vdots    &         & \vdots \\
\delta_y^{n-1}  I_\nu(-y) & \cdots & \delta_y^{2n-2}I_\nu(-y) 
\end{array} \right | .
\label{partffv}
\ee
This expression can be easily rewritten in terms of one of the  
familiar forms for the finite volume partition function.

\subsubsection{The Toda Lattice Equation}
\label{sec:2ptToda}

From the structure of $Z^{(\nu)}_{m,n}(x,y)$ in (\ref{partf}) 
it automatically follows that $Z^{(\nu)}_{m,n}(x,y)$ 
satisfies a Toda
lattice equation. Below, this will be shown by a slight extension of an
argument given in \cite{ForresterBook}.  \\

For the determinant of a matrix $A$ we have the Sylvester idendity
\cite{Sylvester} 
\be
C_{ij} C_{pq} - C_{iq}C_{pj} = \det(A) C_{ij,pq}
\ee 
where $C_{ij}$ is the cofactor of matrix element $ij$ 
\be
C_{ij}\equiv \frac{\del \det(A)}{\del A_{ij}}
\ee
and $C_{ij,pq}$ is the double cofactor of matrix elements $ij$ and
$pq$
\be
C_{ij,pq}\equiv \frac{\del^2 \det(A)}{\del A_{ij}\del A_{pq}}.
\ee
We apply this relation to the matrix elements 
$A^{(m,n)}_{m-1,n-1},\, A^{(m,n)}_{m-1,n},\,A^{(m,n)}_{m,n-1}$
and $A^{(m,n)}_{m,n}$ of $A^{(m,n)}$ entering in eq.
(\ref{ZchGUEasTau}). This results in
\be
C_{m-1n-1} C_{mn} - C_{m-1n}C_{mn-1} = \det(A^{(m,n)}) C_{m-1n-1,mn}.
\ee
Using the explicit form of the derivatives in the determinant of
(\ref{partf}) the cofactors can be expressed as derivatives 
\be
C_{mn} &=& \det A^{(m-1,n-1)},\nn\\ 
C_{m-1n} &=& -\delta_x \det A^{(m-1,n-1)},\nn\\ 
C_{m-1n-1} &=& \delta_x \delta_y \det A^{(m-1,n-1)},\\ 
C_{mn-1} &=& -\delta_y \det A^{(m-1,n-1)},\nn\\  
C_{m-1n-1,mn} &=&  \det A^{(m-2,n-2)}.\nn
\ee
This results in the recursion relation 
\be
 [\delta_x \delta_y \det A^{(m-1,n-1)}] \det A^{(m-1,n-1)}
 -[\delta_x \det A^{(m-1,n-1)}][\delta_y \det A^{(m-1,n-1)}]
= \det A^{(m,n)} \det A^{(m-2,n-2)}.
\ee
Raising the indices of the recursion relation by one this relation
can be rewritten as
\be
\delta_x \delta_y \log \det(A^{(m,n)}) =
\frac{\det A^{(m+1,n+1)} \det A^{(m-1,n-1)}}{   {\det}^2 A^{(m,n)} }.
\ee
The Toda lattice
equation for the generating functional of the two-point function
of the chUE follows immediately 
by applying this identity to (\ref{ZchGUEasTau}). It is given by
\be
\delta_x \delta_y \log Z^{(\nu)}_{m,n}(x,y) = 4nmx^2 y^2
\frac{ Z^{(\nu)}_{m+1,n+1}(x,y) Z^{(\nu)}_{m-1,n-1}(x,y)}
     {[Z^{(\nu)}_{m,n}(x,y)]^2 }.
\label{TodaZtpfCHGUE}
\ee

\subsubsection{The Replica Limit of the Toda Lattice Equation}

The microscopic disconnected scalar susceptibility 
for the chUE is given by 
\be
\chi^{(\nu)}(x,y)\equiv\lim_{m\to 0, n\to 0} \frac{1}{nm}\frac d{dx}
\frac d{dy} \log Z^{(\nu)}_{m,n}(x,y). 
\ee
Using the Toda lattice equation above in this definition we find
\be
\chi^{(\nu)}(x,y)= 4xy Z^{(\nu)}_{1,1}(x,y)Z^{(\nu)}_{-1,-1}(x,y),
\label{twopointchGUE}
\ee
where, explicitly,
\be
Z^{(\nu)}_{1,1}(x,y) = \frac1{y^2-x^2}(x I_{\nu+1}(x)I_\nu(y) 
- y I_{\nu+1}(y)I_\nu(x)),
\ee
and \cite{DV1}
\be 
Z^{(\nu)}_{-1,-1}(x,y) = \frac1{y^2-x^2}(x K_{\nu+1}(x)K_\nu(y) 
- y K_{\nu+1}(y)K_\nu(x)).
\ee
The two point spectral correlation function is given by the
discontinuity of the disconnected scalar susceptibility across the
imaginary axis. The result is in exact agreement with the analytical 
result for the two-point function found in \cite{TV-two}. Again, the 
replica limit of the Toda lattice equation explains that the spectral
correlation function comes from the product of a fermionic and a bosonic
partition function. 

\section{Toda Lattice Equation for Non-Hermitian Theories} 

In this section, we  derive, in the limit of weak non-Hermiticity, 
the spectral density for 
non-Hermitian random matrix theories  from the
replica limit of the Toda lattice equation. We will consider the 
weak non-Hermiticity limit of two systems:
{\sl 1)} A disordered system in an imaginary vector potential (the
Hatano-Nelson model \cite{HN}) which is in the universality class
of the  unitary ensemble. 
We will derive the known spectral density of this system \cite{FKS,E}
from the replica limit of the Toda lattice equation. 
{\sl 2)} QCD at non zero baryon chemical potential which belongs to the
universality class of the chiral  
unitary ensemble. We will                    
derive the quenched microscopic spectral density of the non-Hermitian Dirac
operator. This is a new result that we have checked against a 
high statistics numerical simulation. For a general discussion
of non-Hermitian random matrix theories, we refer to a recent
review by Fyodorov and Sommers \cite{FS-review}.

\subsection{Disordered System at Nonzero Imaginary Vector Potential}
\label{sec:EfetovCase}

\vspace{4mm} 

The spectral density of the Hamiltonian for a disordered system in an 
imaginary vector potential was derived in \cite{FKS,E} using the
supersymmetric method. In \cite{FKS} this result was obtained starting
from an ensemble of complex random matrices, whereas in \cite{E} the starting
point was a nonlinear $\sigma$-model which is also applicable beyond the
ergodic domain.
This nonlinear $\sigma$ model  is applicable to 
the Hatano-Nelson model \cite{HN} and belongs to the universality
class of the (almost Hermitian) UE. 
In this section we analyze its partition function 
       for $n$ fermionic flavors. We find that it
satisfies a Toda lattice
equation enabling us to derive the known result \cite{FKS,E} for the spectral
density by means of the replica trick. 
The replicated bosonic partition function that enters in
the replica limit of the Toda lattice equation is evaluated using
the Ingham-Siegel integral \cite{Yan}. 

\subsubsection{The fermionic partition function}

The static limit of the partition function for $n$ fermionic flavors is
given by \cite{E} (see also the discussion in
\cite{altland-simons}) 
\be
\ZGUE_n(y;a) = \frac 1{{\rm vol}(U(n))}\int dQ 
e^{-\frac {a^2N}{4}{\rm Tr} [Q,\Sigma_3]^2 + yN {\rm Tr}\Sigma_3 Q},
\label{ZnE}
\ee
where the integral is over $U(2n)/[U(n) \times U(n)]$. 
The matrix $\Sigma_3$ is defined by
\be
\Sigma_3 = \mat {\bf 1}& 0 \\  0 & -{\bf 1} \emat
\ee
and ${\bf 1}$ is the $n \times n$ unit matrix. The parameter $a$
determines the strength of the imaginary vector potential,
while $y$ is the mass of the fermionic flavors (i.e. the imaginary
part of the argument of the resolvent). 
In order to evaluate the integral we choose the parameterization
\cite{critique,lerner} 
\be
Q =T^\dagger \Sigma_3 T,
\ee
with 
\be
T = V^\dagger
\mat \cos \theta & \sin \theta \\ -\sin \theta & \cos \theta \emat
V, 
\ee
and  
\be
V =\mat v_1^\dagger & 0 \\ 0 & v_2^\dagger \emat .
\ee
Here, $v_1 \in U(n)$ and $v_2 \in U(n) /U^n(1)$
and $\theta = {\rm diag}(\theta_1,\cdots, \theta_n)$. The integration
measure is given by the product of the matrix elements of the 
off-diagonal blocks of
\be
\delta T \equiv V dT T^{-1} V^\dagger.
\ee
The calculation of the Jacobian of the transformation from the variables
$\{\delta T_{12}, \, \delta T_{21}\}$ to the variables $\{dV V^\dagger,
\delta \cos\theta\}$ is elementary and is given by
\be
J = \prod_k 2\cos \theta_k \prod_{k<l} (\cos^2\theta_k -\cos^2\theta_l)^2.
\ee
The angular integrations factorize from the partition function and
result in the volume of the integration manifold
\be 
V_n = {\rm vol}^2( U(n))/{\rm vol}^n(U(1)).
\ee
In terms of the variables 
\be
\lambda_k = \cos^2\theta_k -\sin^2\theta_k,
\ee
the partition function can thus be written as
\be 
\ZGUE_n(y;a) = 
\frac{V_n}{2^{n(n-1)-n} n!{\rm vol}(U(n)) } 
\int_{-1}^1 \prod_k {\rm d}\lambda_k \prod_{k<l} 
(\lambda_k -\lambda_l)^2
e^{-2a^2N \sum_k (\lambda_k^2 -1) + 2yN \sum_k \lambda_k}.
\ee
This partition function can be written as a determinant of
derivatives
\be
\ZGUE_n(y;a) =
\frac{V_n} {2^{2n(n-1)} N^{n(n-1)} {\rm vol}(U(n)) }  
\det[\partial_y^{p+q} \ZGUE_1(y;a)]_{p,q=0,\ldots,n-1} 
\label{ZEfasTau}
\ee
where 
\be 
\ZGUE_1(y;a) = 2\int_{-1}^1 {\rm d} \lambda e^{-2a^2N(\lambda^2-1)+2yN\lambda}. 
\label{Z1Ea}
\ee 
The partition function (\ref{ZnE}) therefore satisfies the Toda lattice equation (cf. the
discussion in section \ref{sec:2ptToda})
\be
\frac 1{N^2}\partial_y^2 \log \ZGUE_n(y;a) = \frac {8n}{\pi}  
\frac {\ZGUE_{n+1}(y;a)
   \ZGUE_{n-1}(y;a)}{\ZGUE_n^2(y;a)}.  
\label{TodaZE}
\ee

\subsubsection{The Spectral Density}

The spectral density is given by ($z\equiv x+iy$)
\be 
\rho(x,y;a) = 
\lim_{n\to 0} \frac 1{n} \frac 1\pi \frac {d}{dz}\frac {d}{dz^*} 
\log \ZGUE_n(z,z^*;a).
\ee
Taking the replica limit of the Toda lattice equation we thus find 
\be
\rho(x,y;a) = \frac {8N^2} {\pi^2}\ZGUE_1(y;a) \ZGUE_{-1}(y;a) 
\label{rhoE}
\ee
where (cf. (\ref{Z1Ea}))
\be
\ZGUE_{1}(y;a) = 4 e^{2a^2N} 
\int_0^1\cosh(2yNt)\exp\left(-2a^2Nt^2\right){\rm d} t 
\label{Z1Eb}
\ee
and, as we will show below,
\be
\ZGUE_{-1}(y;a) = 
\frac {C_{-1}}a \sqrt{\frac {\pi}{2N}}\exp(-a^2N)
\exp\left(-\frac{Ny^2}{2a^2}\right). 
\label{Zm1E}
\ee
The spectral density is $x$-independent. This implies that if we change
the non-Hermiticity parameter $a$ there is only spectral flow perpendicular
to the $y$-axis. 
Therefore, the normalization $C_{-1}$ of $\ZGUE_{-1}$ has to be chosen
such that the integral 
\be
\frac {8N^2} {\pi^2} \int_{-\infty}^\infty 
dy \ZGUE_1(y;a) \ZGUE_{-1}(y;a) = \frac {32 NC_{-1}}{\pi }e^{Na^2}  
\label{intnorm}
\ee
is independent of $a$. Normalizing the integrated spectral density to
$N$ per unit length we find
\be
C_{-1} = \frac {\pi }{32} e^{-Na^2}.  
\ee
Combining the previous five equations, we find the spectral density 
\be 
\rho(x,y;a) = \frac{N\sqrt N}{\sqrt{2\pi} a}
\exp\left(-\frac{Ny^2}{2a^2}\right) 
\int_0^1\cosh(2yNt)\exp\left(-2a^2Nt^2\right){\rm d} t ,
\ee
in complete agreement with earlier work \cite{FKS,E}. Not only do we
obtain the exact analytical result by means of the  
replica limit, but we have also shown why the spectral density
factorizes into a bosonic and a fermionic partition function.

Before we derive $\ZGUE_{-1}$ let us comment on the $a$-dependence of
the normalization integral (\ref{intnorm}).
The partition function $\ZGUE_{1}$ 
was derived from the effective partition
function (\ref{ZnE}). The commutator term arises because of the requirement
of local gauge invariance \cite{GL,KST,E,KSTVZ} of the effective 
Lagrangian with the kinetic term included. The non-Hermitian random matrix
theory in \cite{FKS} does not have this property. Starting from this
random matrix model \cite{FKS}, which will be introduced in the next section, 
the exponential prefactor in (\ref{Z1Eb}) would have been $\exp(Na^2)$ instead
of $\exp(2Na^2)$ and the normalization integral would have been automatically
$a$-independent.

 
\subsubsection{Calculation of $\ZGUE_{-1}$ using the Ingham-Siegel integral}
\label{sec:Zm1Efetov}

In this subsection we calculate $\ZGUE_{-1}$ 
for a non-Hermitian random matrix model 
using the Ingham-Siegel integral
as proposed in \cite{Yan}. The saddle point manifold is obtained
automatically as a result of this integral. This should be contrasted with
the standard approach which uses the Hubbard-Stratonovitch transformation and
where it is necessary to deform the integration 
contour in order 
to be able to interchange integrations \cite{Schaefer-Wegner}. 
 
The partition function $\ZGUE_{-1}$ is defined by
\be 
\ZGUE_{-1}(z,z^*;a) = \lim_{\epsilon \to 0, N\to\infty} 
C_\epsilon\left\langle {\det}^{-1} \mat \epsilon & z+H+A \\ z^* + 
H - A & \epsilon \emat \right\rangle,
\label{zef-2}
\ee 
where the average $\langle \cdots \rangle$ is over the probability
distribution  
\be
P(H,A) = e^{ -\frac N2 {\rm Tr}H^\dagger H + \frac N{2a^2}{\rm Tr}
 A^\dagger A}, 
\ee
and the $N\times N$ matrices $H$ and $A$ are Hermitian,
$H^\dagger = H$, and anti-Hermitian, $A^\dagger = - A$, 
respectively. 
The normalization constant will be chosen such that the limit $\epsilon
\to 0 $ is finite. In order  to write the
inverse determinant as a convergent bosonic integral
\be
 {\det}^{-1} \mat \epsilon & z+H+A \\ z^* + 
H - A & \epsilon \emat 
=
\int d\phi_i e^{i[\epsilon \phi_1^{*\,k}\phi_1^k+\epsilon\phi_2^{*\,k}\phi_2^k
+\phi_{1}^{*\,k}(z\delta_{kl}+H_{kl}+A_{kl})\phi_{2}^l
+\phi_{2}^{*\,k}(z^*\delta_{kl}+H_{kl}-A_{kl})\phi_{1}^l]},
\ee
the imaginary part of $\epsilon$ has to be positive.
We collect the bosonic fields into 
the positive definite $2\times 2$ matrix  
\be
\bar Q_{ij} \equiv \sum_{k=1}^N (\phi_i^*)^k \phi_j^k.
\ee
If we introduce a mass matrix by 
($\sigma_i$ are the Pauli matrices) 
\be
\zeta = x \sigma_1 - y \sigma_2
\ee
the partition function can be written as
\be 
\ZGUE_{-1}(z,z^*;a) =\lim_{
\epsilon \to 0, N\to\infty}C_\epsilon
\int d\phi_i e^{i{\rm Tr}( (\zeta^T+\epsilon) \bar Q)
-\frac 1{2N} {\rm Tr} \bar Q \sigma_1 \bar Q \sigma_1 - 
\frac {a^2}{2N} {\rm Tr} \bar Q \sigma_2 \bar Q \sigma_2 }.
\ee
The four-boson terms can be linearized by means of a $\delta$ function, 
\cite{hacken,hackenplus} 
\be 
\delta(Q-\bar Q) = \int dF e^{i{\rm Tr} F(Q-\bar Q)}.
\ee
Here both $F$ and $Q$ are Hermitian. This results in 
\be
\ZGUE_{-1}(z,z^*;a) =\lim_{\epsilon\to 0, N\to\infty}C_\epsilon\int d\phi_i dQ dF e^{i {\Tr}(-F+\epsilon)\bar
  Q+i{\Tr}(FQ)+i{\Tr}(\zeta^T Q)
-\frac 1{2N} {\Tr} Q \sigma_1 Q \sigma_1 
-\frac {a^2}{2N} {\Tr} Q \sigma_2 Q \sigma_2 }.
\ee
Under the assumption that the integrals can be interchanged we obtain
\be
\ZGUE_{-1}(z,z^*;a) =\lim_{\epsilon \to 0, N\to\infty}C_\epsilon\int dQ dF {\det}^{-N}(F-\epsilon)
e^{i{\Tr} FQ+{\Tr}( \zeta^T Q)
-\frac 1{2N} {\Tr} Q \sigma_1 Q \sigma_1 - 
\frac {a^2}{2N} {\Tr} Q \sigma_2 Q \sigma_2}.
\ee
The integral over $F$ is an Ingham-Siegel integral of the second kind.
For ${\rm Im}(\epsilon) < 0$ it is given by \cite{Yan}
\be
\int dF {\det}^{-N} (\epsilon + F)  e^{i{\rm Tr}Q F}=
C_{N,p} \theta (Q){\det}^{N-p}(Q) e^{-i\epsilon\Tr Q},
\label{IS}
\ee 
where the integral is over  $p \times p$ Hermitian matrices, 
$C_{N,p}$ is an irrelevant  constant, 
and $\theta(Q)$ denotes that $Q$ is positive definite.
After rescaling $Q \to N Q$ we find ($N>2$)
\be
\ZGUE_{-1}(z,z^*;a) =\lim_{\epsilon \to 0, N\to\infty}C_\epsilon
\int dQ \theta(Q) {\det}^{N-2}Q 
e^{iN \,{\Tr}((x\sigma_1+y\sigma_2+\epsilon) Q)
-\frac N2 {\Tr} Q \sigma_1 Q \sigma_1 - 
\frac{a^2 N}2 {\Tr} Q \sigma_2 Q \sigma_2 }.
\label{ZmnCalcText} 
\ee 
We calculate this integral in the weak non-Hermiticity limit and
the microscopic limit, i.e. $N\to \infty$ with
$\alpha^2 N$ and $z N$ fixed. In this limit, the saddle point equation, 
given by
\be
(\sigma_1 Q)^2 = 1,
\label{saddleZGUEm1} 
\ee 
fixes two of the four degrees of freedom in $Q$. In order to
complete       the evaluation of $\ZGUE_{-1}$, we will choose an explicit
parametrization of $Q$ and perform the integration over the two 
degrees of freedom not fixed by the saddle point equation exactly. 
Positive definite 
$2\times2$ Hermitian matrices can be parameterized as
\be
Q = e^t \mat e^u \cosh s  & i e^{i\phi}\sinh s \\
          -i e^{-i\phi}\sinh s &   e^{-u} \cosh s \emat.
\ee
The Jacobian for the transformation from the $Q$ variables to
the variables of this parameterization is given by
\be 
J = 4 e^{4t}\cosh s \sinh s.
\ee
The saddle point condition (\ref{saddleZGUEm1}) is
given by 
\be
t =0, \qquad \phi = 0,
\ee
whereas the variable $s$ and $u$ are not determined by the saddle point
equations. They parameterize the saddle point manifold.
The Gaussian fluctuations in the $t$ and $\phi$ 
variables about this saddle point manifold give rise
to a factor $1/(2\sinh s)$. 
As final result we thus obtain 
\be
\ZGUE_{-1}(z,z^*;a) = \lim_{\epsilon \to 0, N\to \infty}
 2C_\epsilon\int_{-\infty}^\infty du         
\int_{-\infty}^\infty ds \cosh \,s\, e^{a^2 N(1 -2\cosh^2 s)
-i2Ny \sinh s +2i N\epsilon \cosh u\cosh s}.
\ee 
The integral over $u$ results in a contribution $\sim-\log\epsilon$. 
This leading order singularity in $\epsilon$ will be canceled by the
normalization constant $C_\epsilon$ and the remaining prefactors, 
$2C_\epsilon$, will 
be denoted by $C_{-1}$.
The integral over $\sinh s$ can then be performed by completing squares. This 
results in the Gaussian $y$ dependence (\ref{Zm1E})
\be
\ZGUE_{-1}(y;a) = 
\frac {C_{-1}}a \sqrt{\frac {\pi}{2N}}\exp(-a^2N)
\exp\left(-\frac{Ny^2}{2a^2}\right). 
\label{Zm1Eb}
\ee

The Toda lattice equation suggests that the partition function with 
$n$ bosonic flavors can be expressed as a determinant of derivatives. 
This leads us to conjecture
\be
\ZGUE_{-n}(y;a) & = & C_{-n}
\det[\partial_y^{p+q}\ZGUE_{-1}(y;a)]_{p,q=0,\ldots,n-1}
\label{ZEbasTauConj}\\ 
& \sim &   \frac{\exp[-\frac{n N y^2}{2a^2}]}{a^{n^2}}
\ee
where $\ZGUE_{-1}$ is given by (\ref{Zm1E}). 
This conjecture will be discussed in a separate 
publication \cite{FSV}.

\subsection{The Phase Quenched QCD Partition Function}
\label{sec:phqchGUE}

In this section we derive the microscopic spectral density of 
the quenched Euclidean
QCD Dirac operator at nonzero baryon chemical potential. The calculation
is performed in the weak non-Hermiticity limit, where only terms to up
to second order in the chemical potential contribute to the effective
action. In this subsection we follow the usual convention that the
massless Dirac operator at zero chemical potential is anti-Hermitian. Therefore
``weak non-Hermiticity'' is actually ``weak non-anti-Hermiticity'', but we 
will refrain from using this name.  

The generating function for the quenched spectral density at nonzero 
chemical potential is not the QCD partition function with $n$ flavors,
but rather the QCD partition function with $n$ flavors and $n$ conjugate
flavors \cite{Girko,misha}. Since the fermion determinants of the flavors and 
the conjugate flavors are each others complex conjugate this partition function
is also  known as the phase quenched QCD partition function.
For $n$ flavors with mass $z$ and $n$ flavors with mass $z^*$ it is given by
(in the sector of zero topological charge)
\be
Z_n(z,z^*;\mu) = \left
                \langle {\det}^n\mat z & id+\mu \\ id^\dagger+\mu & z \emat
                     {\det}^n\mat z^* & -id+\mu \\ -id^\dagger +\mu &
                                                                z^* \emat  
             \right\rangle_{\rm YM},\qquad z = x+ iy,
\label{pqQCD}
\ee
where $id$ is the covariant derivative and the average over gauge
field configurations is weighted by the Yang-Mills action.
For positive $n$ 
the  low-energy limit of this theory in the chirally broken
phase is \cite{dominique-JV}
\be
Z_n(z,z^*;\mu) = \frac 1{(2\pi)^{n(2n+1)} \, {\rm vol}(U(n))}
\int_{U \in U(2n)} dU e^{-\frac {VF_\pi^2\mu^2}4
{\rm Tr} [U,B][U^\dagger,B] + \frac 12 \Sigma V {\rm Tr}(MU + MU^\dagger)}
\label{zeff}
\ee
with\footnote{We stress that the combination $(MU + MU^\dagger)$ enters
the action and not the usual combination $(M^\dagger U + MU^\dagger)$. 
The reason is that the phase quenched QCD action (\ref{pqQCD})
has $n$ flavors of mass $z$ and other $n$ flavors has mass
$z^*$. The mass term that follows from chiral invariance
is given by $\Tr(M_{RL}U+M_{LR}U^\dagger)$ with
 $M_{RL}$  the right-left handed mass matrix and $M_{LR}$ the
left-right handed mass matrix. The usual mass term is obtained
if $M_{RL} = M_{LR}^\dagger$. We however have that
$M_{RL}=M_{LR} = M$ with  $M$ complex.} 
\be
B = \Sigma_3 \qquad {\rm and} \qquad M = \mat z& 0 \\ 0 & z^* \emat.
\ee
The volume of space-time is denoted by $V$.
The resolvent is defined by
\be
G(z,z^*;\mu) = \lim_{n\to0}\frac 1{nV} \partial_z \log Z_n(z,z^*;\mu).
\ee
At nonzero $\mu$ the support of the spectrum of the Dirac operator 
is a two-dimensional domain in the complex plane with spectral density
given by
\be
\rho(z;\mu) =\frac 1\pi \partial_{z^*} G(z,z^*;\mu).
\ee
In the thermodynamic limit at fixed $\mu$ and $m$, the effective partition
function (\ref{zeff}) can be analyzed in terms of mean field theory
\cite{KSTVZ,dominique-JV,Misha-Son}. The result is \cite{dominique-JV}
that the spectral density
is constant inside the domain
\be
|{\rm Re} z| < \frac {2\mu^2F^2}\Sigma
\label{meanfield}
\ee 
and zero outside this domain.

\subsubsection{Toda Lattice Equation}

Here we will analyze the effective partition function in the 
microscopic limit where both $\mu^2 V$ and $ zV$ remain fixed in the
thermodynamic limit. Our aim is to calculate the spectral density
of the quenched theory. 
To do this we will show that $Z_n$ satisfies the Toda lattice
equation. The proof follows directly from  the proof for 
two point function at zero 
chemical potential: In the parameterization discussed in section
\ref{sec:2pf} the trace of the first term in the action is given by
\be
 {\rm Tr} [U,B][U^\dagger,B] = 8\sum_{k=1}^n (\lambda_k^2 -1).
\ee

Repeating the steps (\ref{limit})-(\ref{partf}) with this factor included we
find (we recall the notation $\delta_z\equiv z \partial_z$) 
\be
Z_n(z,z^*;\mu) =\frac{D_n}{(z z^*)^{n(n-1)}} \det \left[\delta_z^k
  \delta_{z^*}^l Z_1(z,z^*;\mu)\right]_{k,l=0,1,\ldots,n-1}, 
\label{ZpqchGUEasTau}
\ee
with 
\be 
Z_1(z,z^*;\mu) = \frac 1\pi e^{2VF^2_\pi\mu^2}\int_0^1 d\lambda \lambda 
e^{-2VF^2_\pi\mu^2\lambda^2} I_0(\lambda z\Sigma V)I_0 (\lambda z^*\Sigma V)
\label{z1quenched}
\ee
and
\be 
D_n = \frac{2^{n(n+1)/2}}{\pi^{n(n-1)/2}\prod_{k=1}^{n-1}k!}.
\ee
This results in the Toda lattice equation 
\be 
\delta_z \delta_{z^*} \log Z_n(z,z^*;\mu) = \frac{\pi n}2(zz^*)^2\frac
       {Z_{n+1}(z,z^*;\mu)Z_{n-1}(z,z^*;\mu)}{Z_n^2(z,z^*;\mu)}. 
\label{TodaPQchGUE} 
\ee 
The quenched spectral density is then given by
\be
\rho(x,y;\mu) &=& \lim_{n\to 0} \frac 1{\pi n} \partial_z\partial_{z^*} 
\log Z_n(z,z^*;\mu)                     
\nn \\ &=&\lim_{n\to 0} \frac { z z^*}{2} 
\frac       {Z_{n+1}(z,z^*;\mu)Z_{n-1}(z,z^*;\mu)}{Z_n^2(z,z^*;\mu)}.
\label{replicaLimTodaPQchGUE}
\ee
In our normalization $Z_0(z, z^*) = 1$ so that
\be
\rho(x,y;\mu) &=&  \frac {z z^*}{2} Z_{1}(z,z^*;\mu)Z_{-1}(z,z^*;\mu).
\label{rhochGUE} 
\ee
Our next task is to calculate $Z_{-1}(z,z^*;\mu)$. To do this we derive
the static low energy partition function from a chiral random matrix model
corresponding to (\ref{pqQCD}).  The main problem is to choose
the correct integration manifold. As is discussed 
in the next section, this can be done without guess work if we
use the Ingham-Siegel integral as proposed in \cite{Yan}. As an introduction
to this method we have included the
calculation of $Z_{-1}(z,z^*)$ for $\mu = 0$
in \ref{App:Zm1mu=0}. 

\subsubsection{The effective partition function $Z_{-1}$}
 
In this section we derive the effective phase quenched partition
function in the microscopic and weak non-Hermiticity limit. 
The integration manifold is determined by symmetries and
convergence requirements. Usually, these conditions define a
suitable integration contour for the $\sigma$-field.
A method that avoids this  somewhat ad hoc procedure was advocated in
\cite{hacken,Yan}. Below we will follow this approach in the calculation of
the phase quenched partition function for one bosonic flavor.
 
The partition function for one bosonic flavor and one conjugate bosonic flavor
is given by the integral 
\be 
\lim_{N\to \infty}
\int dW e^{-\frac{N}{2}{\rm Tr} W W^\dagger} 
{\det}^{-1} \mat z& iW + \mu\\iW^\dagger +\mu &z \emat   
{\det}^{-1} \mat z^*& -iW + \mu\\-iW^\dagger +\mu &z^* \emat. 
\label{zchguemu}
\ee
If $z$ is inside the domain of eigenvalues, the partition function
potentially diverges. Therefore, we have to regularize the determinants.
A suitable regularization is one where
the determinant can be written as a convergent bosonic integral. This 
procedure, which is also known as hermitization \cite{FZ}, amounts to rewriting
the product of the two determinants as
\be
&&{\det}^{-1} \mat z& iW + \mu\\iW^\dagger +\mu &z \emat   
{\det}^{-1} \mat z^*& -iW + \mu\\-iW^\dagger +\mu &z^* \emat  \\
\label{detzm1}
& = & \lim_{\epsilon \to 0}
{\det}^{-1} \matf \epsilon &       0  &               z &    iW+\mu\\
      0                  & \epsilon & iW^\dagger +\mu & z \\
      z^*                &-iW + \mu & \epsilon        & 0 \\
      -iW^\dagger+\mu        & z^*      & 0               & \epsilon \ematf.
\nn
\ee 
After rearranging the rows and columns inside the determinant, the 
partition function (\ref{zchguemu}) can be written as
\be
Z_{-1} &=&\lim_{\epsilon \to 0, N\to \infty}C_\epsilon\int dW 
e^{-\frac{N}{2}{\rm Tr} W W^\dagger} 
 \int d\phi_kd\phi_k^*
\exp[ i  
\vect \phi_1^*\\\phi_2^*\\\phi_3^*\\\phi_4^*\evect^T 
\matf \epsilon & z        &  iW+\mu & 0\\
      z^*   &\epsilon   & 0& iW - \mu         \\
    -iW^\dagger +\mu&  0               & \epsilon &-z^*\\
      0 &  -iW^\dagger -\mu &-z& \epsilon  \ematf
\vect \phi_1\\ \phi_2\\\phi_3\\\phi_4\evect] \nn. \\
\label{detzm2}
\ee 
After averaging over $W$ the limit $\epsilon \to 0$ is potentially singular.
Such singularity will be absorbed as a multiplicative renormalization 
constant into $C_{\epsilon}$. 
As in the case of the Hatano-Nelson model the overall
prefactor of $Z_{-1}$ 
will be fixed by the normalization of the complex
density. For this reason we do not keep explicitly track of the prefactor,
it will be simply labeled by $C_\epsilon$.

The Gaussian integration over $W$ results in a 4-boson interaction term
given by
\be
\exp \left [ -\frac 2N {\rm Tr} \bar Q_1 \bar Q_2\right ],
\ee
where
\be
\bar Q_1 \equiv \mat \phi_1^*\cdot \phi_1 & \phi_1^*\cdot \phi_2\\
                   \phi_2^*\cdot \phi_1 & \phi_2^*\cdot \phi_2 \emat ,
\qquad
\bar Q_2 \equiv \mat \phi_3^*\cdot \phi_3 & \phi_3^*\cdot \phi_4\\
                   \phi_4^*\cdot \phi_3 & \phi_4^*\cdot \phi_4 \emat , 
\label{Q12}
\ee 
and we have used the notation
\be
\phi^*_k \cdot \phi_l = \sum_{i=1}^{N/2} \phi_k^{i\,*} \phi_l^i.
\ee 
Instead of the usual Hubbard-Stratonovitch transformation, we linearize
the 4-boson interaction term with the help of matrix $\delta$ functions.
For Hermitian matrices $Q_i$ and $\bar Q_i$ the $\delta$-function 
can be represented as
\be
\delta(Q_i -\bar Q_i) = \frac 1{(2\pi)^4}
\int dF e^{-i {\rm Tr}F(Q_i-\bar Q_i)} , 
\label{matdelta}
\ee 
where the integral is over Hermitian matrices $F$. This results in
\be
\exp \left [-\frac 2N {\rm Tr} \bar Q_1  \bar Q_2  \right ]
= \frac {1}{(2\pi)^8}
\int dQ_1dQ_2 \int dF dG e^{ {\rm Tr} [-i F(Q_1 -\bar Q_1)
-iG(Q_2 -\bar Q_2) -\frac 2N Q_1  Q_2  ]}. 
\ee
The integral over the $\phi_k$ is uniformly convergent in  
$F$ and $G$ which allows 
us to interchange the order of these integrals.
We finally obtain the partition function
\be
Z_{-1} = \lim_{\epsilon\to0,N\to \infty}C_\epsilon 
\int dQ_1dQ_2 \int dF dG e^{ {\rm Tr }[-i FQ_1 - 
iGQ_2 -\frac  2N Q_1 Q_2 ]} {\det}^{-\frac{N}{2}}
\mat \zeta + F^T & \mu\sigma_3 \\ \mu\sigma_3 & 
-I\zeta I + G^T \emat, 
\ee
where we have used a block notation and
\be
\zeta = \mat \epsilon & z \\ z^* & \epsilon \emat.
\label{masszeta}
\ee
The anti-symmetric matrix $I$ is defined by
\be
I = \mat 0 & 1 \\ -1 & 0 \emat.
\ee
We further simplify this integral by changing integration variables according
to
$F \to F-\zeta^T$ and $G\to G+I \zeta^TI$ and $Q_i \to N Q_i/2,
\, i= 1,2 $. This results in
\be
Z_{-1} = \lim_{\epsilon \to 0,N \to\infty}C_\epsilon 
\int dQ_1dQ_2 \int dF dG e^{ {\rm Tr} [-i \frac{N}{2}FQ_1 -
i\frac{N}{2}GQ_2 +i \frac{N}{2}\zeta^T(Q_1 -IQ_2 I) -\frac{N}{2}  
Q_1  Q_2  ]} {\det}^{-\frac{N}{2}}
\mat \epsilon  + F & \mu\sigma_3 \\ 
\mu\sigma_3 & \epsilon + G  \emat,
\ee
where, for reasons of convergence we also left  infinitesimal increments
inside the determinant.
Contrary to
the example in \ref{App:Zm1mu=0}, we were not able to perform
the $F$ and $G$ integrations analytically. However, in the weak
non-Hermiticity limit, 
where $\mu^2N$ remains fixed in the limit $N\to \infty$, 
the determinant is given by  
\be 
{\det}^{-\frac{N}{2}}
\mat \epsilon  +F & \mu\sigma_3 \\ \mu\sigma_3 & \epsilon + G \emat
= {\det}^{-\frac{N}{2}} (\epsilon +F){\det}^{-\frac{N}{2}} (\epsilon + G)
\exp\left [\frac{N\mu^2}{2} {\rm Tr}(\frac 1{\epsilon +F}\sigma_3 
\frac 1 {\epsilon + G} \sigma_3)\right ](1 +{\cal O}\left(\frac1N\right)).
 \ee 
In the limit $N\to \infty$, at fixed $\mu^2N$, 
the  $F$ and $G$ variables in the $\mu^2 N$ term
can be replaced with the saddle point values of $F$ and $G$ at
$\mu=0$. They are given by 
\be
\frac 1{\epsilon + F} = i  Q_1, \qquad 
\frac 1{\epsilon + G} = i  Q_2.
\ee
The remaining integrals over $F$ and $G$ are Ingham-Siegel
integrals (see (\ref{IS})). They can be performed exactly 
resulting in
\be 
Z_{-1}(z,z^*;\mu) = \lim_{\epsilon \to 0, N\to\infty}
C_\epsilon
\int dQ_1dQ_2 \theta(Q_1) \theta(Q_2) {\det}^{\frac{N}{2}-2}(Q_1Q_2)  
e^{ {\rm Tr }[
 i \frac{N}{2}\zeta^T(Q_1 -IQ_2I) -\frac{N}{2} Q_1  Q_2   
-\frac{N}{2}\mu^2 Q_1 \sigma_3 Q_2 \sigma_3 ]}, 
\ee 
where we remind the reader that $\zeta$ contains the regulator mass $\epsilon$.  
(Recall that $\theta(Q)$ is
a matrix step function which is equal to unity  if $Q$ is Hermitian and 
positive definite and equal to zero otherwise). 
In the limit $N\to \infty$, the integrals over the massive modes can
be performed by a saddle point approximation.
The saddle point equations are given by
\be
Q_1^{-1}-Q_2 = 0, \qquad  Q_2^{-1} -Q_1 = 0.
\ee
Both equations can be rewritten as
\be
Q_1 = Q_2^{-1},
\ee
and therefore only four of the modes, which we choose to be $Q_2$, 
can be integrated out by a 
saddle point approximation. The quadratic fluctuations give rise to a factor
$\pi^2/{\det}^2 Q_1$.
The integral over the remaining modes has
to be performed exactly. We thus arrive at the partition function
\be
Z_{-1}(z,z^*;\mu) = 
\lim_{\epsilon\to0} C_\epsilon
\int \frac{dQ_1}{{\det}^2 Q_1} \theta(Q_1) 
e^{ {\rm Tr} [ 
 i \frac{N}{2}\zeta^T(Q_1 -I Q_1^{-1}I )   
-\frac{N}{2}\mu^2 Q_1 \sigma_3Q_1^{-1} \sigma_3 ]}. 
\label{ZMINQ}
\ee 
Comparing this result with the fermionic effective partition function
(\ref{zeff}) we can make the following identification
\be
N &\longleftrightarrow& V,\nonumber \\
\Sigma &\longleftrightarrow& 1,\label{identification} \\
\mu^2 &\longleftrightarrow& \mu^2 F_\pi^2. \nonumber
\ee 
We will evaluate this integral explicitly in section
\ref{subsec:Zm1chGUEExplicit}. Before doing so we now rederive this
partition function based on the symmetries of the QCD partition function.

\subsubsection{Symmetries of $Z_{-1}$}

In this section we show that the effective partition function (\ref{ZMINQ}) 
is completely determined by the symmetries of 
the underlying microscopic partition function. 
This implies that any theory with the same pattern of  global symmetries
and spontaneous symmetry breaking is described by the same low-energy
effective theory. The only memory of the microscopic theory are 
the two constants, $\Sigma$ and $F_\pi$, in (\ref{ZMINQ}). In particular,
this means that the low-energy limit of QCD  and the random matrix model
discussed in the previous section are the same.
The virtue of the random matrix approach is that 
the low-energy limit of the partition function can be derived directly
from the microscopic theory.

For $ \zeta =0$ and $\mu = 0$ the random matrix 
partition function (\ref{detzm2}) is invariant under
$Gl(2)/U(2) \times U(2)$. Explicitly, the axial $Gl(2)/U(2)$ invariance is
realized by
\be
\vect \phi_1\\ \phi_2\\\phi_3\\\phi_4\evect 
\to 
\mat U_A & 0 \\ 0 & U^{-1}_A \emat
\vect \phi_1\\ \phi_2\\\phi_3\\\phi_4\evect \qquad
{\rm with} \qquad
U_A^\dagger U_A^{-1} = 1.
\label{axial}
\ee
For reasons of convergence we can allow only symmetry transformations
that do not alter the complex conjugation structure of the partition function.
Therefore,
\be
U_A = e^H, \quad {\rm with} \quad H^\dagger = H.
\ee
The vector $U(2)$ symmetry is realized by
\be
\vect \phi_1\\ \phi_2\\\phi_3\\\phi_4\evect 
\to 
\mat U_V & 0 \\ 0 & U_V \emat
\vect \phi_1\\ \phi_2\\\phi_3\\\phi_4\evect \qquad
{\rm with}\qquad
U_V^\dagger U_V = 1.
\label{vector}
\ee
Imposing these symmetries at nonzero mass and chemical potential 
requires that, in the block notation,
\be
\mat \zeta_1 & \mu_1 \\ \mu_2 &\zeta_2 \emat
\ee
the mass and chemical potential are transformed as 
\be
\zeta_1 \to U_A^{-1} \zeta_1 U_{A}^{-1}, & \qquad &
\mu_1 \to U_A^{-1} \mu_1 U_A,\nn\\
\zeta_2\to U_A \zeta_2 U_{A} \ \ \ \, , & \qquad &
\mu_2\to U_A \mu_2 U_A^{-1},
\label{vectormass}
\ee
under axial transformations and as
\be
\zeta_1 \to U_V \zeta_1 U_{V}^{-1}, & \qquad &
\mu_1 \to U_V \mu_1 U_V^{-1},\nn\\
\zeta_2 \to U_V \zeta_2 U_{V}^{-1}, & \qquad &
\mu_2 \to U_V \mu_2 U_V^{-1}
\label{axialmass}
\ee
under vector transformations.
The matrix $\bar Q_1$ introduced in (\ref{Q12}) transforms as
\be
\bar Q_1 \to U_A^T \bar Q_1 U_A^T, \qquad \bar Q_1 \to U_V^* \bar Q_1 U_V^T,
\label{Qtrans}
\ee
under axial and vector transformations, respectively. 
To first order in the mass matrix and second order in the chemical potential,
we therefore can write down the following nontrivial invariants ($Q_1$ has
the same transformation properties as $\bar Q_1$)
\be
{\rm Tr} Q_1 \zeta_1^T, \qquad
{\rm Tr} Q_1^{-1} \zeta_2^T, \qquad
{\rm Tr} Q_1 \mu_2^TQ_1^{-1} \mu_1^T.
\ee
Using that
\be
\zeta_1 = \zeta, \quad \zeta_2 = -I \zeta I, \quad \mu_1 =\mu_2 =\mu\sigma_3,
\ee
we find the nontrivial invariant terms
\be
 {\rm Tr} Q_1 \zeta^T, \qquad
-{\rm Tr} Q_1^{-1} I\zeta^TI, \qquad
\mu^2{\rm Tr} Q_1 \sigma_3 Q_1^{-1} \sigma_3.
\ee
The effective partition function should be  invariant
under the discrete symmetry, $\zeta \to -I \zeta I$, 
of the microscopic partition function as well. 
Because the integration measure on positive definite Hermitian matrices,
$dQ_1 /{\det}^2 Q_1$, is invariant under $Q_1 \to Q_1^{-1}$ this is
the case if the mass terms occur in the combination
\be
{\rm Tr} Q_1 \zeta^T - {\rm Tr} Q_1^{-1} I\zeta^TI.
\label{massterm}
\ee
The numerical prefactors of the invariant terms are not fixed by the
global symmetries of the random matrix model but follow from matching 
to the microscopic theory.

At  nonzero chemical potential the phase quenched QCD partition function 
with one pair of bosonic flavors has to be hermiticized 
as in the random matrix model (\ref{detzm1}) 
resulting in a Dirac operator with the structure
given in (\ref{detzm1}) and a mass matrix as in (\ref{masszeta}).
The partition function has the global axial and vector symmetries of 
(\ref{axial}) and (\ref{vector}) if the mass and chemical potential matrices
are
transformed according to (\ref{vectormass}) and (\ref{axialmass}) as well as
the discrete symmetry $\zeta \to -I\zeta I$. 
The Goldstone manifold is parameterized
by the axial tranformations and is therefore
the coset of positive definite matrices \cite{OTV,DOTV} which transform
according to (\ref{Qtrans}). At zero chemical potential, the 
effective Lagrangian density to order $p^2$ respecting these symmetries 
is therefore given by 
\be
L_{-1}= \frac{F_\pi^2}4 {\rm Tr} \partial_\mu Q \partial_\mu Q^{-1}
-i\Sigma \frac 12({\rm Tr} Q \zeta^T - {\rm Tr} Q^{-1} I\zeta^TI).
\ee
The phase of the mass term is determined by the condition that the integral
over $Q$ is convergent for $\epsilon >0$.
For real quark masses, $z=z^*=m$,
the mass term is minimized by the saddle-point
solution $Q = -i\sigma_1$. The vacuum energy is therefore given by
$2m\Sigma V $ which identifies $\Sigma$ as the chiral condensate. As is
the case in the supersymmetric formulation of partially quenched chiral
perturbation theory \cite{BG}, 
the pion decay constant in $L_{-1}$ is taken to be the same as in
the case of fermionic quarks.

The chemical potential enters in the QCD partition function as an external
vector field,
\be
V_\nu = \mu \sigma_3 \delta_{\nu 0}.  
\label{vectormu}
\ee
Therefore, the vector symmetry (\ref{vector}) can be promoted
to a local symmetry by transforming $V_\nu$ as a nonabelian
gauge field
\be
V_\nu \to U_V^{-1}\partial_\nu U_V  + U_V^{-1} V_\nu U_V.
\ee
The effective Lagrangian at nonzero chemical potential 
should respect this symmetry. This is achieved by replacing the derivative
in the kinetic term by the covariant derivative
\be
\del_\nu Q \to \nabla_\nu Q \equiv \del_\nu Q + [V_\nu, Q]
\ee
with $V_\nu$ given by (\ref{vectormu}).
The low energy effective Lagrangian density is therefore given by 
\be
L_{-1}= \frac{F_\pi^2}4 {\rm Tr} \nabla_\nu Q \nabla_\nu Q^{-1}
-i\Sigma \frac 12({\rm Tr} Q \zeta^T - {\rm Tr} Q^{-1} I\zeta^TI).
\ee
In a parameter domain where the fluctuations of the zero momentum modes
are much larger than the fluctuations of the nonzero momentum modes
the zero momentum part factorizes from the partition function \cite{GLeps} 
and is given by
\be
Z_{-1}(z,z^*;\mu) = 
\lim_{\epsilon\to0} C_\epsilon
\int \frac{dQ}{{\det}^2 Q} \theta(Q) 
e^{ {\rm Tr} \left[ 
 i \frac{V}{2}\zeta^T(Q -I Q^{-1}I )   
-\frac{V}{4}F_\pi^2 \mu^2 [Q ,\sigma_3][Q^{-1} ,\sigma_3 ]\right]} ,
\label{ZMINQCD}
\ee 
where $dQ/{\det}^2 Q \theta(Q)$ is the integration measure on positive
definite Hermitian matrices. Up to an overall constant, this result 
agrees  with  the effective random matrix partition function
imposing the identification (\ref{identification}).

\subsubsection{Evaluation of $Z_{-1}$}
\label{subsec:Zm1chGUEExplicit}

We parameterize the positive definite matrix $Q_1$ in (\ref{ZMINQ}) as
\be
Q_1 = e^t \mat e^r \cosh s  & e^{i\theta}\sinh s \\
          e^{-i\theta}\sinh s &   e^{-r} \cosh s \emat.
\ee 
The Jacobian of this transformation is easily found to be 
\be
J = 4 e^{4t}\cosh s \sinh s.
\ee
The range of the integration variables is given by 
\be
r \in \langle -\infty, \infty \rangle, \quad
s \in \langle -\infty, \infty \rangle, \quad
t \in \langle -\infty, \infty \rangle, \quad
\theta \in \langle 0, \pi \rangle \quad.
\ee
Inserting this parameterization 
in (\ref{ZMINQ}) we find that the partition function is given
by the following integral
\be
Z_{-1} &=& 
\lim_{\epsilon\to0}C_\epsilon 
\int dr ds dt d\theta
\cosh s |\sinh s| 
e^{i\frac{N}{2}(4x\sinh s \cosh t \cos \theta - 4 y \sinh s \sinh t
  \sin \theta +4\epsilon \cosh r \cosh s \cosh t) -\mu^2 N (1+2 \sinh^2 s)}.
\ee
The integral over $r$ results in the  modified Bessel 
$K_0(2N\epsilon \cosh s \cosh t)$ with leading singularity given by 
$\sim -\log \epsilon$. This factor is absorbed in the normalization of 
the partition function. 
Then the integral over $\theta$ gives a
Bessel function. Introducing $u = \sinh s$ as new integration variable
we obtain
\be 
Z_{-1}(z,z^*;\mu) &=&
C_{-1}\int_{-\infty}^\infty dt  \int_0^\infty du u J_0(2N u(x^2 \cosh^2 t
+y^2 \sinh^2 t)^{1/2})
e^{-\mu^2 N (1+2 u^2)},
\label{zminj}
\ee
where the finite prefactor has been labeled $C_{-1}$. 
The integral over $u$ is a known integral over a Bessel function. 
The final result for the phase quenched partition function
with one bosonic flavor is thus given by
\be
Z_{-1}(z,z^*;\mu) &=& \frac {C_{-1}e^{-N\mu^2} }{4\mu^2N } 
\int_{-\infty}^\infty  dt  
e^{-\frac{N (x^2 \cosh^2 t +y^2 \sinh^2 t)}{2\mu^2}}\nn \\
& =& 
\frac {C_{-1}e^{-N\mu^2} }{4\mu^2N }
e^{\frac{N (y^2-x^2)}{4\mu^2}} K_0(\frac{N (x^2+y^2)}{4\mu^2}).\label{zm1final}
\ee

\subsubsection{The Spectral Density in the Complex Plane}

Reminding that, with the identification (\ref{identification}), the partition
function for one fermionic flavor is given by
\be 
Z_1(z,z^*;\mu) = \frac 1\pi e^{2N\mu^2}\int_0^1 d\lambda \lambda 
e^{-2N\mu^2\lambda^2} I_0(\lambda zN)I_0 (\lambda z^*N),
\ee
the final result for the spectral density of the chGUE in the weak
non-Hermitian limit obtained from the replica limit
(\ref{replicaLimTodaPQchGUE}) of the Toda lattice equation
(\ref{TodaPQchGUE}) is given by ($z=x+iy$) 
\be
\rho(x,y;\mu) &=&  \frac {z z^*}2 Z_{1}(z,z^*;\mu)Z_{-1}(z,z^*;\mu) \nn \\
 &=&  \frac {C_{-1}e^{N\mu^2}}{8\pi\mu^2 N}(x^2+y^2)
e^{\frac{N (y^2-x^2)}{4\mu^2}} K_0(\frac{N (x^2+y^2)}{4\mu^2})
\int_0^1 d\lambda \lambda  e^{-2N\mu^2\lambda^2} I_0(\lambda zN)I_0
(\lambda z^*N). 
\label{rhochGUEfinal}
\ee
As in section (\ref{sec:EfetovCase}) the constant $C_{-1}$ can 
be obtained from a spectral flow argument.
In this case, the spectral
flow with variation of $\mu$ 
becomes perpendicular to the $y$ axis only for $Ny^2 \gg \mu$.
In this limit we can use the large argument asymptotic approximation of
the Bessel functions,
\be
K_0(z) = \sqrt{\frac \pi{2z}} e^{-z},\quad 
I_0(z) = \frac 1{\sqrt{2\pi z}} ( e^z +i e^{-z}),\quad 
I_0(z^*) = \frac 1{\sqrt{2\pi z^*}} ( e^{z^*} -i e^{-z^*}),
\ee
resulting in the large $y$ asymptotic limit of the spectral density
(note that the oscillating terms in $(z-z^*)N\lambda$ are subleading after
integration over $\lambda$)
\be 
\rho(x,y;\mu) 
 &=&  \frac {C_{-1}e^{N\mu^2}}{4(2\pi)^{3/2} N^{5/2} \mu}
e^{\frac{-Nx^2}{2\mu^2}} 
\int_0^1 d\lambda \lambda  e^{-2N\mu^2\lambda^2} 
(e^{2\lambda Nx} + e^{-2\lambda N x}).
\label{rhoasym}
\ee
The integral over $x$ of this expression can be calculated analytically,
\be
\int_{-\infty}^\infty \rho(x,Ny^2\gg\mu^2;\mu)dx =
\frac{C_{-1}e^{N\mu^2}}{4\pi N^3}
\ee
and should be independent of $\mu$. 
In the normalization of $N$ eigenvalues per unit length in the $y$ direction
this results in 
\be
C_{-1} =  4\pi N^4 e^{-N\mu^2}.
\label{normcm1}
\ee
The microscopic spectral density, obtained in the limit $N\to \infty$ 
with $\mu_s = \mu^2 N$ and $\eta = z N$ fixed, 
is thus given by
\be
\rho_s(\eta,\eta^*;\mu_s) &=& \lim_{N\to \infty} \frac 1{N^2} 
\rho(\frac{{\rm Re}(\eta)}N,\frac{{\rm Im}(\eta)}N, 
\frac{\mu_s}{\sqrt N}) \nn \\
& = &
 \frac {\eta\eta^*}{2\mu^2_s}
\exp(-\frac{\eta^2+\eta^{*\,2}}{8\mu^2_s}) K_0(\frac{\eta\eta^*}{4\mu^2_s})
\int_0^1 d\lambda \lambda  e^{-2\mu^2_s\lambda^2} I_0(\lambda \eta)I_0
(\lambda \eta^*). 
\label{finalmicro}
\ee
Again we note that $Z_{-1}$ was calculated starting from a chiral random 
matrix model, whereas $Z_1$ was obtained from an effective partition
function with a commutator term that was a remnant of the local gauge
invariance of the chiral Lagrangian. If we also would have derived $Z_{1}$  
from the random matrix model, the exponential prefactor would have been
$e^{N\mu^2}$ instead of $e^{2N\mu^2}$, and the normalization integral
would have been automatically $\mu$-independent.

The leading order asymptotic expansion of the
bosonic partition function in $\mu^2/N(x^2+y^2)$ is given by
\be
Z_{-1}(x,y;\mu) =\frac {C_{-1}\sqrt \pi e^{-N\mu^2}}
                       {2\mu N \sqrt{2N(x^2+y^2)} }
e^{-\frac{N x^2}{2\mu^2}}(1+{\cal O}\left(\frac{\mu^2}{N(x^2+y^2)}\right)). 
\ee
If we insert this result in the expression for 
the spectral density (\ref{rhochGUEfinal}) we reproduce the 
result in \cite{gernotSpectra} which was obtained by means of the
complex orthogonal method \cite{FS-complex-polynoms} for a closely 
related partition function defined in an eigenvalue representation. 
The discrepancy between the phase quenched partition function
(\ref{zeff}) and the partition function in \cite{gernotSpectra}
is even more remarkable since it
was shown \cite{gernotUniversal} that, in versions of both models with
the phase of the fermion determinant not quenched, the two models coincide 
in the weak non-Hermiticity limit.

\begin{figure}[h!]
\vspace{7cm}
 \unitlength1.0cm
  \begin{center}
\begin{picture}(3.0,2.0)
  \put(0,0){
  \psfig{file=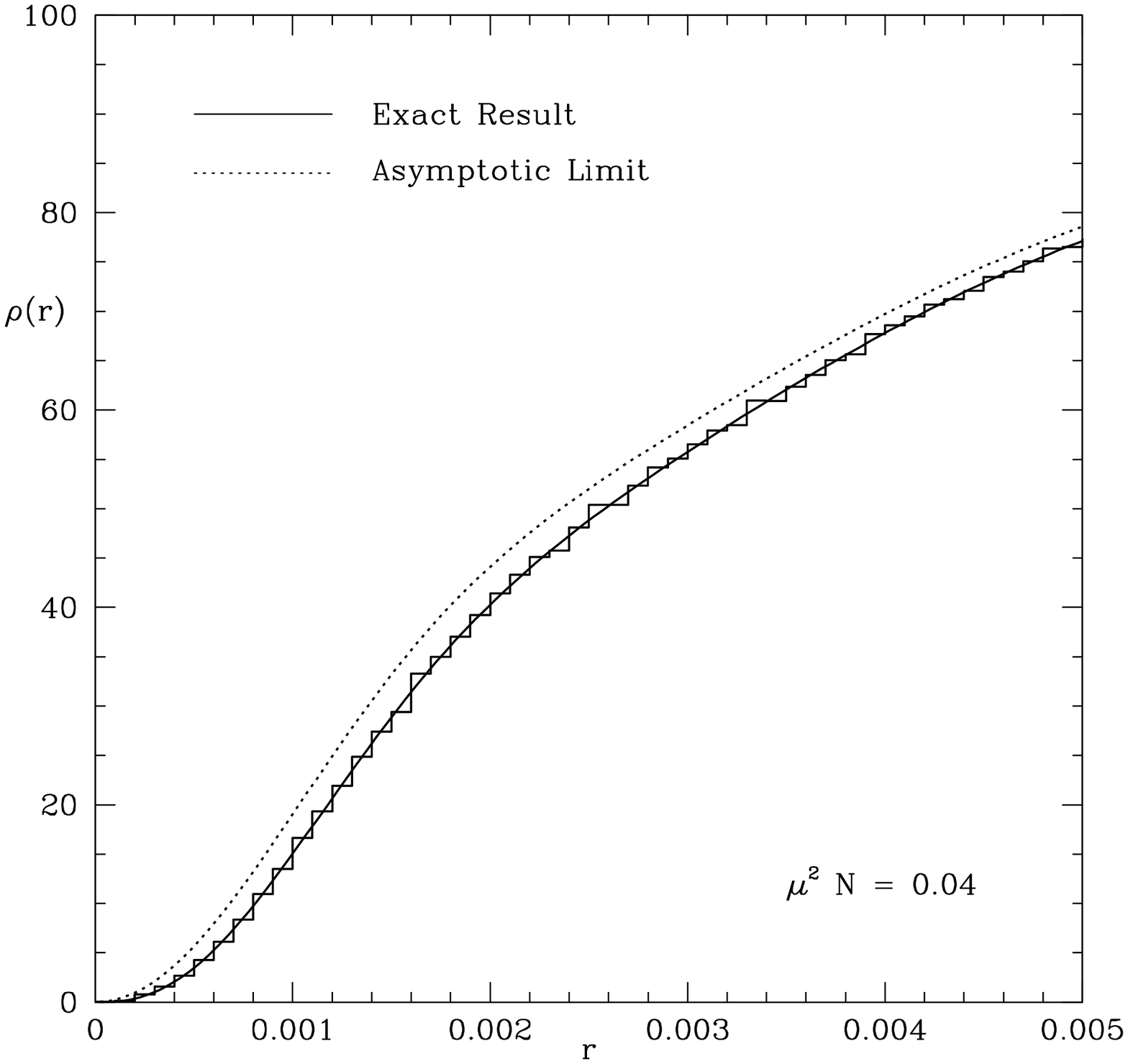,clip=,width=9.0cm}}
\end{picture}
\hfill
  \begin{picture}(3.0,2.0)
  \put(-6.0,0){
  \psfig{file=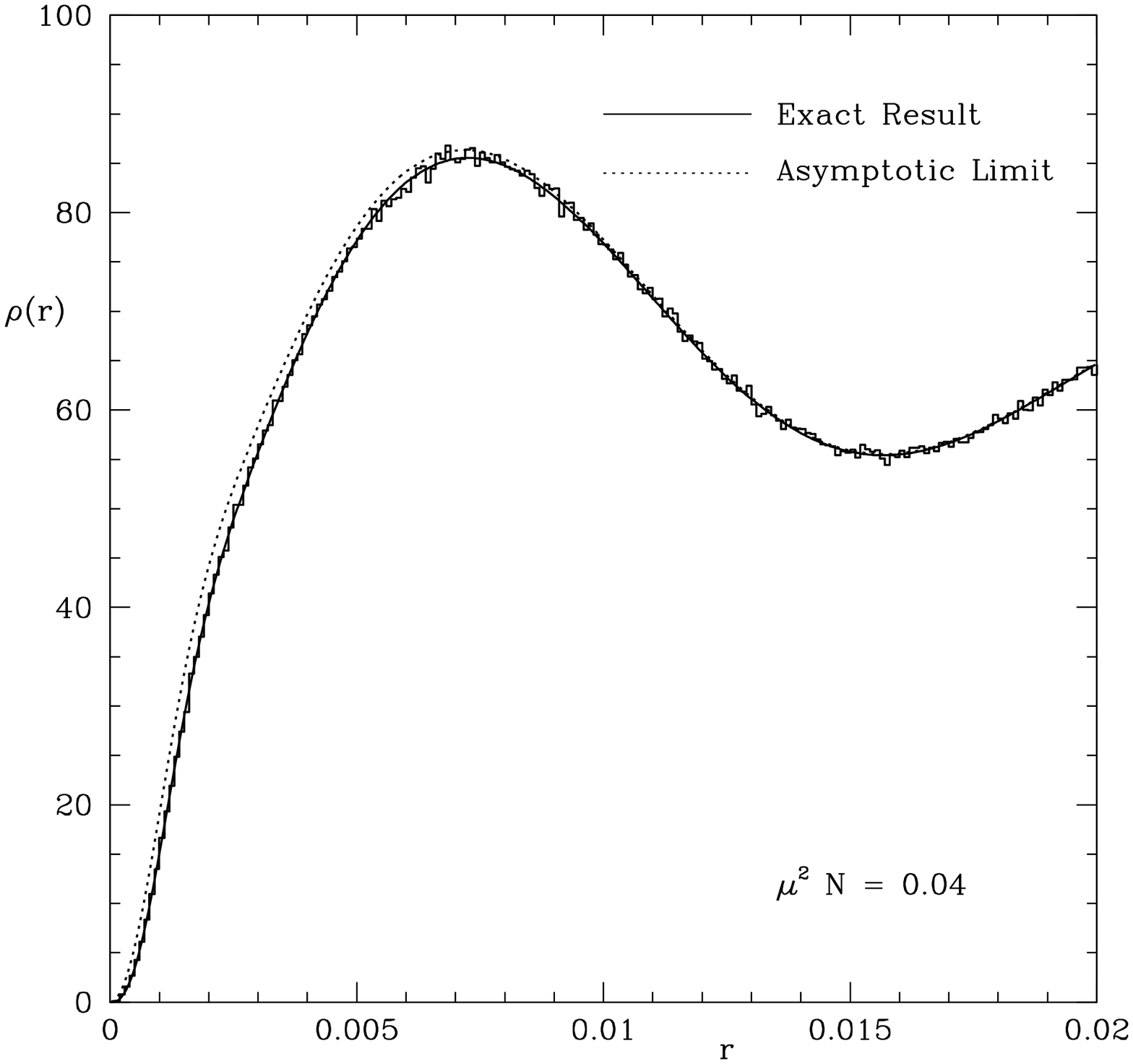,clip=,width=9.cm}}
  \end{picture}
  \end{center}

\caption{The radial microscopic 
  spectral density of the quenched QCD partition function 
  at nonzero chemical potential.
  The histogram shows the
  result of a numerical simulation in which 2.000.000 random matrices
  of size $200 \times 200$ where diagonalized. 
  The value of the chemical potential is given by 
  $\mu^2N=0.04$ with $N=200$. The full line is the exact analytical
  result and is in agreement with the data. Also shown is the result with
  the asymptotic limit of the bosonic partition 
  function in the expression for the density (dotted curve). This curve
fails to describe the data when
  $r/\mu$ is not in the asymptotic domain $r/\mu\gg 1$. The left figure
is a blown-up version of the right figure for $r < 0.005$.}
\label{Fig}   
\end{figure}

To convince the reader of the correctness of our result, (\ref{finalmicro}),
we have performed a numerical diagonalization of 2000000 matrices of
the form
\be
\left(\begin{array}{cc} 0 & iW+\mu \\ iW^\dagger+\mu & 0
\end{array}\right).
\ee
The matrices where of size $N=200$ and drawn on the weight (\ref{weightW}).
In Fig. \ref{Fig}, we have plotted a histogram of the radial density,
\be
\rho(r;\mu) =\int_0^{2\pi} d\phi\rho(x=r\cos\phi,y=r\sin\phi ;\mu)
\ee
for $\mu^2 N = 0.04$. In the same figure we plot our analytical result 
(solid curve) and the leading order asymptotic approximation of our result
(dotted curve) which coincides with the result of \cite{gernotSpectra}.  
Recently, Dirac spectra were calculated for  
quenched lattice QCD at nonzero chemical potential \cite{GT}. Although the
lattice 
spectra 
were shown \cite{GT} to be in agreement with 
\cite{gernotSpectra}, the statistical
accuracy is not sufficient to distinguish between the 
two analytical 
results. We are looking
forward to simulations that will resolve this issue.

In the limit $y ^2 N \gg \mu^2$ and $\mu^2 N \gg 1$ the eigenvalues are
distributed homogeneously in a strip parallel to the $y$-axis as has been shown
by a mean field argument \cite{dominique-JV}. In this limit  
the Bessel functions in (\ref{finalmicro}) can be approximated by their
asymptotic expansion (see (\ref{rhoasym})) resulting in 
\be 
\rho_s(\eta,\eta^*;\mu_s)
& = & 
 \frac {1}{2\mu_s\sqrt{2\pi}}
\exp(-\frac{(\eta+\eta^{*})^2}{8\mu^2_s}) 
\int_0^1 d\lambda e^{-2\mu^2_s\lambda^2}(e^{\lambda(\eta+\eta^*)}
+e^{-\lambda(\eta+\eta^*)}). 
\label{limitmicro}
\ee
The saddle point of the $\lambda$-integral is located at
\be
\bar \lambda = \frac{\eta+\eta^*}{4\mu_s^2}.
\ee
If this saddle point is outside the integration domain, the integral 
is exponentially suppressed. We thus find that in the limit $\mu_s \gg 1$
and ${\rm Im}(\eta) \gg \mu^2_s$
the spectral density vanishes for
\be 
\frac{|\eta+\eta^*|}{4\mu_s^2} = \frac {|x|}{2\mu^2} >1.
\ee
If the saddle point is inside the integration domain, a saddle point
approximation gives
\be  
\rho_s(\eta,\eta^*;\mu_s)
& = & \frac 1{4\mu_s^2} \qquad {\rm for} \qquad
\frac{|\eta+\eta^*|}{4\mu_s^2} = \frac{|x|}{2\mu^2} <1,
\ee 
in agreement with the mean field result (\ref{meanfield}) 
\cite{dominique-JV} provided that  
we make the identification (\ref{identification}).
The total number of eigenvalues in a strip of unit width perpendicular to
the $y$-axis is thus given by  
\be
N^2 \rho_s(\eta,\eta^*;\mu_s) 4\mu^2  = N,
\ee
in agreement with the choice of our normalization constant in
(\ref{normcm1}).

Finally, let us stress that we expect that it is possible to express
$Z_{-n}$ as a determinant of matrix with elements given by derivatives of
$Z_{-1}$. This leads us to conjecture the following form of $Z_{-n}$
\be
Z_{-n}(z,z^*;\mu) =\frac{C_{-n}}{(z z^*)^{n(n-1)}} \det \left[\delta_z^k
  \delta_{z^*}^l Z_{-1}(z,z^*;\mu)\right]_{k,l=0,1,\ldots,n-1}. 
\label{ZmnpqchGUEasTau}
\ee
We hope to prove this conjecture in a future publication \cite{FSV}.


\section{Conclusions and Outlook}
\label{sec:conc}

We have shown that the replica limit of the Toda lattice equation is a
powerful tool to derive the spectral correlation function of both Hermitian
and non-Hermitian random matrix theories. We have obtained
one-point and two-point
functions of Hermitian random matrix ensembles in the class A (Wigner-Dyson)
and in class AIII (chiral) and one-point functions for non-Hermitian
random matrix theories also both in class A and in class AIII. 
In the case of the chiral ensemble, the non-Hermiticity arose
because of a chemical potential that was introduced in the random matrix model
consistent with the covariance properties of the QCD partition function.
Therefore, this result is particularly relevant for the analysis of 
Dirac spectra of quenched QCD at nonzero chemical potential. 
We emphasize 
that our analytical results for this case were not previously known.
These results convincingly show that it is possible to derive nonperturbative
results by means of the replica trick contradicting the lore that the method
only can be trusted for the calculation of asymptotic series 
of analytical expressions. 
In addition, the replica limit of the Toda lattice equation explains the
factorization of correlation functions 
in terms of a product of a fermionic and a bosonic partition function. 
Such a structure is present both in Hermitian and non-Hermitian random
matrix theories. 

Although the results for the correlation functions in the other cases 
we have considered were known previously several new insights have emerged.
In particular, we have conjectured that 
the fermionic, the bosonic and the super-symmetric partition
functions are $\tau$-functions of a single integrable hierarchy which
are related by means of the Toda lattice equations. In fact, this is the
basis for the success of the replica trick in this approach. We have shown
the validity of this conjecture for the fermionic 
generating function of the chGUE two-point function and the fermionic
generating functions of the non-Hermitian theories. In several other cases
this property was known previously. In the bosonic case, only
the generating function for the GUE two-point function 
was shown to  satisfy the fermionic Toda lattice equation for an arbitrary
number of replicas. 
Of course, in cases  were the spectral correlation
functions were known previously it is clear that the Toda lattice equation
can be continued at least to the partition function with one bosonic flavor. 
The difficulty with bosonic partition functions is that 
the weight in the integral has poles due to the inverse determinant. 
For Hermitian random matrix theories
the poles are on the real axis and can be easily avoided. In the case
of non-Hermitian random matrix theories with eigenvalues scattered in the
complex plane, the poles are located in the very domain where we wish to
calculate the spectral density. These problems can be dealt with by
hermiticizing the generating function and using the Ingham-Siegel
integral to avoid ambiguities in the choice of the integration contours.
Although we have only evaluated the bosonic partition function for one
bosonic flavor, we do not expect major technical problems for an arbitrary
number of flavors and we expect to find the result that we have conjectured
based on the Toda lattice equation. This problem will be addressed in a
future publication.    
 
Recently, quenched lattice QCD Dirac spectra at nonzero chemical potential
were compared to a non-Hermitian chiral random matrix model that was formulated
in terms of eigenvalues so that the method of complex orthogonal polynomials
could be applied to this model. 
The spectral density of this model 
differs from our result, but since the leading order asymptotic expansion of the
two models is the same, the two results are close and, within the
statistical accuracy, they are both in agreement with the  
existing 
lattice data. However, in a numerical 
simulation of a chiral random matrix model at nonzero chemical potential, we
convincingly have shown 
that the model of Akemann disagrees with the data, whereas
our result is right on the mark. Based on universality arguments one 
would expect that the two models would coincide in the weak non-Hermiticity
limit, but apparently, the model of Akemann is in a different universality
class. 

Having shown that it is possible to obtain exact analytical results 
by means of the replica trick some comments on the critique of the
replica trick set forward in \cite{critique,zirn} are in order.
One of the points that were raised is that the fermionic partition
function or the bosonic partition functions alone are not sufficient
to reproduce an exact analytical result. This is particularly clear
in the fermionic case where the resolvent is an analytic function of
the mass for all positive integer values of the replica index, so that its
discontinuity is zero even in the replica limit. A careful analysis 
\cite{zirn} shows that one cannot expect to obtain the full analytical
answer in the bosonic case either. 
The conclusion of \cite{zirn} was that,
in order to derive nonperturbative results, the proper procedure is
to combine information from bosonic and fermionic replicas. This is
exactly what is achieved by the replica limit of the Toda lattice equation.
However, the Toda lattice equation itself does not justify an analytical
continuation from integer values of the number of replicated flavors. Only in
the case of the one point function of the unitary ensemble with finite size
matrices is it known how to derive a Toda lattice equation for any real
number of flavors. 

It would be of great conceptual and practical interest to generalize
the results obtained in this paper to the orthogonal and symplectic ensembles. 
The orthogonal chiral ensemble is relevant for QCD with two colors where 
lattice simulations at nonzero chemical potential already have been 
performed. In this case there is no sign problem and one can go beyond 
the quenched approximation. Several predictions for macroscopic properties 
\cite{KST,KSTVZ,dominique-JV,eff} have been confirmed by means of lattice gauge
simulations \cite{latt}. A comparison to the microscopic spectral 
correlation functions will offer an independent crosscheck of both 
numerical and analytical methods.

\vspace{1cm}

\noindent
{\bf Acknowledgments:} 
This work was supported in part by 
U.S. DOE Grant No. DE-FG-88ER40388.
We wish to thank the ECT$^*$ in Trento where
 part of this work was completed. Also our thanks extends to
Gernot Akemann, Poul Damgaard, Yan Fyodorov, Eugene Kanzieper and Martin Zirnbauer 
for useful discussions and suggestions.

\vspace*{1cm}\noindent

\renewcommand{\thesection}{Appendix \Alph{section}}
\setcounter{section}{0}

\section{Notation} 
\label{App:Notation}

In this appendix we comment on the notation used in this paper. 
In the main text we consider several the generating functions or partition
functions. A partition
function  in the class of the Unitary Ensemble
(abbreviated UE) will be denoted by $\ZGUE$  and  a partition function
in the class of the chiral Unitary Ensemble
(abbreviated chUE) by $Z$. If the probability distribution of the matrix
elements of matrices in these ensembles is Gaussian, they will be
called the Gaussian Unitary Ensemble (GUE) and the chiral Gaussian Unitary
Ensemble (chGUE), respectively. If the form of the partition function does
not depend on the specific form of the probability distribution we often
omit the Gaussian adjective. 
The subscripts are used in the same way for both
ensembles. We denote the partition function for $m$ flavors of
mass $x$ and $n$ flavors of mass $y$ by $\ZGUE_{m,n}(x,y)$ and
$Z^{(\nu)}_{m,n}(x,y)$. The additional superscript $\nu$ of the
chUE partition function denotes the topological sector. This superscript will
be dropped for  the trivial topological sector (i.e., $\nu=0$). 
The sign of negative subscripts will be written out explicitly. It indicates
that the corresponding flavors obey bosonic statistics. 
At nonzero imaginary vector potential or chemical potential the partition
function depends both on the complex mass $z$ and it complex conjugate mass
$z^*$. We use the notation $z=x+iy$ and thus have
$\ZGUE_{n}(z,z^*;a)=\ZGUE_{n}(x,y;a)$ and
$Z_{n}^\nu(z,z^*;\mu)=Z_{n}^\nu(x,y;\mu)$, respectively.   
Following the usual convention in the literature, the operators or matrices 
will
be chosen Hermitian or weak non-Hermitian in the GUE class and anti-Hermitian
or weak non-anti-Hermitian in the chUE class. However, for both case we
will use the term ``weak non-Hermiticity'' when appropriate. 

Throughout this paper we will use the unnormalized Haar measure.

\section{Calculation of a Jacobian}
\label{App:Jacobian}

In this Appendix, we calculate the Jacobian of the transformation
\be
U = \mat u_1 & \\ & u_2 \emat \mat v_1 & \\ & v_2 \emat  
\Lambda \mat v_1^\dagger & \\ & v_2^\dagger \emat, \label{param}
\ee
where $U$ is an $(n+m)\times (n+m)$ unitary matrix and  
$\Lambda$ is the block diagonal matrix given by
\be  
\Lambda_{k,l} &=& \mat \sqrt{ 1 - \mu^2 }& \mu\\ \mu & 
-\sqrt{1-\mu^2} \emat_{k,l},\qquad k, l = 1, \cdots, 2m, \nn \\ 
\Lambda_{k,l} &=& 0, \qquad k > 2m \quad {\rm or} \quad l > 2m,  
\quad k \ne l,  
\nn \\ 
\Lambda_{k,k} &=& -1, \qquad k > 2m.
\label{unm-trans} 
\ee
The diagonal matrix $\mu     = {\rm diag}( \mu_1,\cdots, \mu_m)$, 
and we will use the notation
that $\lambda_k = \sqrt{1-\mu_k^2}$ with $\lambda_k\in[0,1]$. Here,
$u_1$ and $v_1$ are unitary $m\times m$ matrices, $u_2 \in U(n)$ and
$v_2 \in U(n) / (U^m(1) \times U(n-m))$.  One easily verifies that the total
number of parameters on both sides of (\ref{param}) is the same. For the 
calculation of the Jacobian of this transformation it is
convenient to use the differentials
\be
\delta U &=& \mat v_1^\dagger & \\ & v_2^\dagger 
\emat U^{-1} dU \mat v_1 & \\ & v_2 \emat, \nn\\
\delta u_1 &=& v_1^\dagger u_1^{-1} du_1 v_1, \nn \\
\delta u_2 &=& v_2^\dagger u_2^{-1} du_2 v_2,\nn \\
\delta v_1 &=& v_1^{-1} dv_1,\nn \\
\delta v_2 &=& v_2^{-1} d v_2\nn \\
\delta \lambda &=& d \lambda.
\ee
The matrices $\delta U$, $\delta u_k$ and $\delta v_k$ are anti-Hermitian. To
calculate the Jacobian we have to distinguish six different cases depending
on whether $\delta U$ is diagonal or either of its indices are smaller or 
larger than
$2m$. We split $\delta U$ into 4 blocks 
\be
\delta U = \mat \delta U_{11} & \delta U_{12} \\
                \delta U_{21} & \delta U_{22}  \emat
\ee
of size $m\times m$, $m \times n$, $n\times m$ and $n\times n$, respectively.

The Jacobian of the transformation from 
$\{\delta U_{11}^{kl}, \delta U_{12}^{kl}, \delta U_{21}^{kl}, 
\delta U_{22}^{kl}\}$ to the variables 
$\{\delta v_1^{kl}, \delta v_2^{kl}, \delta u_1^{kl}, \delta u_2^{kl}, 
\delta \lambda_k\} $ will be denoted by $J_{kl}$.
For $1\le  k \le m$ and $1 \le l \le m$ and $k \ne l$ the Jacobian 
of transformation from  
$\{\delta U_{11}^{kl}, \delta U_{12}^{kl}, \delta U_{21}^{kl}, 
\delta U_{22}^{kl}\}$ to 
$\{\delta v_1^{kl}, \delta v_2^{kl}, \delta u_1^{kl}, \delta u_2^{kl}\} $
is given by the determinant
\be
J_{k\le m, l\le m, k\ne l} &=& \det \matf 
  \lambda_k\lambda_l -1 & \mu_k\mu_l & \lambda_k \lambda_l &\mu_k\mu_l \\
  \lambda_k \mu_l &  -\mu_k \lambda_l & \lambda_k \mu_l &-\mu_k \lambda_l \\
  \mu_k \lambda_l & -\lambda_k \mu_l  & \mu_k \lambda_l & -\lambda_k \mu_l \\
  \mu_k\mu_l & \lambda_k\lambda_l -1 & \mu_k\mu_l & \lambda_k \lambda_l \ematf 
\nn \\ &=& \lambda_l^2 -\lambda_k^2 \label{J1},
\ee
where we have used that $\mu_k^2 +\lambda_k^2 = 1$.
The Jacobian of the transformation from 
$\{ \delta U_{12}^{kl}, \delta U_{22}^{kl}\}$ to the variables 
$\{\delta v_2^{kl},  \delta u_2^{kl}\}$ 
for $1\le  k \le m$ and $m<l \le n$ 
is given by the determinant
\be
J_{k\le m, l > m} &=& \det \mat
  \mu_k & \mu_k \\ -\lambda_k -1 & -\lambda_k \emat \nn \\
&=& \mu_k.
\ee
The Jacobian of the transformation from 
$\{ \delta U_{21}^{kl},\delta U_{22}^{kl}\}$ to the variables 
$\{\delta v_2^{kl},  \delta u_2^{kl}\}$ 
for $k> m$ and $ 1 \le l \le m$ 
is given by the determinant
\be
J_{k> m, l \le m} &=& \det \mat
  \mu_l & \mu_l \\ -\lambda_l -1 & -\lambda_l \emat \nn \\
&=&\mu_l.
\ee
The Jacobian of the transformation from 
$\{ \delta U_{22}^{kl}\}$ to the variables 
$\{\delta u_2^{kl}\}$ 
for $k> m$ and $ l > m$ is simply given by the determinant
 \be
J_{k> m, l > m} &=& 1.
\ee
Finally, we consider the Jacobian of the transformation of the diagonal
matrix elements. 
For the transformation
$\{\delta U_{11}^{kk}, \delta U_{12}^{kk}, \delta U_{21}^{kk}, 
\delta U_{22}^{kk}\}$ to the variables 
$\{\delta v_1^{kk}, \delta u_1^{kk}, \delta u_2^{kk}, 
\delta \lambda_k \}$ with $k <m$ 
we find the Jacobian
\be
J_{k\le m, k\le m} &=& \det \matf 
  -\mu_k^2 & \lambda_k^2 & \mu_k^2 & 0 \\
  \lambda_k \mu_k &  \mu_k \lambda_k & -\lambda_k \mu_k &-1/\mu_k \\
  \mu_k \lambda_k & \lambda_k \mu_k  & -\mu_k \lambda_l & 1/\mu_k \\
  \mu_k^2 & \mu_k^2 &  \lambda_k^2 & 0 \ematf 
\nn \\ &=& 2\lambda_k \label{J5}.
\ee
For $k>m$ the only nonzero  derivatives are 
$\delta U^{22}_{kk}/\delta u^2_{kk} =1$ resulting into a Jacobian of unity.

Multiplying the Jacobians calculated in previous paragraph we find
the Jacobian of the transformation (\ref{param}):
\be
J = \prod_{1\le k<l\le m} (\lambda_k^2 -\lambda_l^2)^2 
\prod_{k=1}^m (2\lambda_k) \mu_k^{2(n-m)}.\label{jactot}
\ee

\section{Calculation of $Z_{-1}$ for $\mu = 0$}
\label{App:Zm1mu=0}

This Appendix illustrates the use of the Ingham-Siegel integral for
the chiral unitary ensemble at $\mu = 0$ 
which was already considered in \cite{Yan}.
In this case there is no need to hermiticize the Dirac operator, and
the inverse determinants can be represented as

\be
&&{\det}^{-1} \mat z& iW \\iW^\dagger &z \emat   
{\det}^{-1} \mat z^*& -iW \\-iW^\dagger &z^* \emat  \nn \\
&=& 
{\det}^{-1} \mat z& iW \\iW^\dagger &z \emat   
{\det}^{-1} \mat z^*& iW \\iW^\dagger &z^* \emat  \nn \\
&=& C\int d\phi_k d\phi_k^* 
\exp[   
-\vect \phi_1^*\\\phi_2^*\evect^T 
\mat  z               & iW\\
       iW^\dagger & z \emat
\vect \phi_1\\\phi_2\evect 
-\vect \phi_3^*\\\phi_4^*\evect^T 
 \mat    z^*      &iW  \\
   iW^\dagger &  z^*     \emat
\vect \phi_3\\\phi_4\evect ].
\ee
Other constants that enter in the calculation will also be absorbed in 
the irrelevant normalization constant $C$. The integrals converge 
if ${\rm Re}(z
) > 0$.
The Gaussian average with probability distribution as in (\ref{zchguemu})
results in
(up to an overall constant)
\be
\exp  \left [ -\frac 2N {\rm Tr} \bar Q_1 \bar Q_2\right ]
\ee
with
\be
\bar Q_1 = \mat \phi_1^*\cdot \phi_1 & \phi_1^*\cdot \phi_3\\
                   \phi_3^*\cdot \phi_1 & \phi_3^*\cdot \phi_3 \emat ,
\qquad
\bar Q_2 = \mat \phi_2^*\cdot \phi_2 & \phi_2^*\cdot \phi_4\\
                   \phi_4^*\cdot \phi_2 & \phi_4^*\cdot \phi_4 \emat . 
\ee
Introducing $\delta$-functions as in (\ref{matdelta}) and performing the 
$\phi_k$-integrations we obtain
\be
Z_{-1} = C
\int dQ_1dQ_2 \int dF dG e^{ {\rm Tr }[i FQ_1 +
iGQ_2 -\zeta Q_1 -\zeta Q_2-\frac  2N Q_1 Q_2 ]} {\det}^{-{N/2}}
(  F ){\det}^{-{N/2}}( G),
\ee
where an infinitesimal imaginary increment is included in $F$ and $G$
and $\zeta $ is now defined by
\be
\zeta = \mat z & 0 \\ 0 & z^* \emat.
\ee
The integrals over $F$ and $G$ can be calculated by means of the 
Ingham-Siegel integral (\ref{IS}). After also rescaling 
$Q_i \to N Q_i/2$ 
we find
\be 
Z_{-1} = C
\int dQ_1dQ_2 \theta(Q_1) \theta(Q_2) e^{ {\rm Tr }[
 -N\frac{\zeta}{2}  Q_1 -N\frac{\zeta}{2} Q_2- \frac N2 Q_1 Q_2 ]} {\det}^{\frac N2-2}
(  Q_1 Q_2 ).
\ee
In the microscopic limit where $\zeta N$ is fixed as $N\to\infty$ the
saddle point equations are given by 
\be 
Q_1 Q_2 = 1.
\ee
We choose the $Q_1$ variables to parameterize the  saddle point manifold. 
The Gaussian fluctuations about $Q_2 = Q_1^{-1}$ give rise to a factor
$1/{\det}^2 (Q_1)$. As final result we obtain
\be 
Z_{-1} = C
\int \frac{dQ_1}{{\det}^2 Q_1} \theta(Q_1)  e^{ {\rm Tr }[
 -N\frac{\zeta}{2}  (Q_1+Q_1^{-1})  ]} .
\ee
We parameterize the positive Hermitian matrices as
\be
Q_1 = U \mat e^{s_1} & 0 \\ 0 & e^{s_2} \emat U^{-1}.
\label{parh}
\ee
The integration measure is given by
\be
\frac{dQ_1}{{\det}^2 Q_1} = (e^{s_1} - e^{s_2})(e^{-s_1} - e^{-s_2})
ds_1 ds_2 dU.
\ee
The integral over $U$ can be calculated as an Itzykson-Zuber integral
\cite{IZ}. This results in
\be 
Z_{-1} = C
\int_{-\infty}^\infty ds_1 \int_{-\infty}^\infty ds_2  
(e^{s_1} - e^{s_2})(e^{-s_1} - e^{-s_2})
\frac{e^{-Nz \cosh s_1 - Nz^*\cosh s_2} -e^{-Nz \cosh s_2 - Nz^*\cosh s_1}}
{N(z-z^*)(\cosh s_1 - \cosh s_2)}.
\label{zdirect}
\ee 
Writing $z = x+ iy$, the limit $y \to 0$ of this partition function is given
by
\be
Z_{-1} = C
\int_{-\infty}^\infty ds_1 \int_{-\infty}^\infty ds_2  
(e^{s_1} - e^{s_2})(e^{-s_1} - e^{-s_2})e^{-Nx (\cosh s_1 +\cosh s_2)}
= 8C(K_0^2(Nx) - K_1^2(Nx)),
\label{zxdirect}
\ee
which, up to an  (arbitrary) normalization constant, agrees with the result
obtained by means of different methods \cite{DV1}.
For $z$ purely imaginary we did not succeed to further simplify the 
integral (\ref{zdirect}).

\section{ The $\mu = 0$ limit of eq. (\ref{ZMINQ})}
\label{App:muto0}

To convince the reader that the partition function with one bosonic
flavor is correct we consider in this Appendix the $\mu = 0$ limit
of (\ref{ZMINQ}) which is given by 
\be
Z_{-1}(z,z^*;\mu) = C
\int \frac{dQ_1}{{\det}^2 Q_1} \theta(Q_1) 
e^{ {\rm Tr} [ 
 i \frac{N}{2}\zeta^T(Q_1 -I Q_1^{-1}I ) ]}.
\label{zminqa}
\ee 
where
\be
I = \mat 0 & 1\\ -1 & 0 \emat \qquad {\rm and}
\qquad    \zeta = \mat 0 & z \\ z^* & 0 \emat.
\ee
As in the previous Appendix, we parametrize positive definite
Hermitian matrices as
\be
Q_1 = U \mat e^{s_1} & 0 \\ 0 & e^{s_2} \emat U^{-1}
\label{parhp}
\ee
with  integration measure given by
\be
\frac{dQ_1}{{\det}^2 Q_1} = (e^{s_1} - e^{s_2})(e^{-s_1} - e^{-s_2})
ds_1 ds_2 dU.
\ee

To rewrite the integral over $U$ as an Itzykson-Zuber integral, we use
the property of $2\times 2$ matrices
\be
I Q^{-1} I = -\frac {Q^T}{\det Q},
\ee
so that
\be
{\rm Tr}\zeta^T(Q_1 -I Q_1^{-1}I )&=&
{\rm Tr}\zeta^T Q_1 + \frac 1{\det Q_1}{\rm Tr} \zeta Q_1 \nn \\
&=& {\rm Tr}(\zeta^T  + \frac 1{\det Q_1} \zeta) Q_1 .
\ee
Evaluating the Itzykson-Zuber integral 
we find as final result   
\be 
Z_{-1} = C
\int_{-\infty}^\infty ds_1 \int_{-\infty}^\infty ds_2  
(e^{s_1} - e^{s_2})(e^{-s_1} - e^{-s_2})
\frac{e^{i\frac{N}{2} (\lambda_1e^{s_1}+ \lambda_2 e^{s_2}) }
-e^{i\frac{N}{2}(\lambda_1 e^{s_2} + \lambda_2 e^{s_1})}}
{i(N/2)(\lambda_1-\lambda_2)(e^{s_1} - e^{ s_2})}.
\label{zmu0}
\ee
where $\lambda_1 $ and $\lambda_2$ are the solutions of
\be 
\lambda^2 = (z + z^* e^{-s_1-s_2})(z^* + z e^{-s_1-s_2}). 
\ee
We did not succeed to show that the partition function (\ref{zmu0}) agrees
with (\ref{zdirect}) for arbitrary complex values of $z$. The reason
is that the integration contour has to be modified appropriately so that
the integral becomes convergent. However, for $z$ purely real or purely
imaginary, (\ref{zdirect}) can be recovered form (\ref{zmu0}).
For $z=iy$ (with $y$ real) one easily shows 
that the integrands of the two partition functions
are the same. This is not the case for $z =x$ (with $x$ real). In this
case a convergent integral is obtained if
we first shift the integration contours in (\ref{zmu0}) by
\be
s_1 \to s_1 +\frac \pi 2 i, \qquad s_2 \to s_2 -\frac \pi 2 i
\ee
in the first term and by
\be
s_1 \to s_1 -\frac \pi 2 i, \qquad s_2 \to s_2 +\frac \pi 2 i
\ee
in the second term. We then find
\be
Z_{-1}(x) = C
\int_{-\infty}^\infty ds_1 \int_{-\infty}^\infty ds_2  
(e^{s_1} + e^{s_2})(e^{-s_1} + e^{-s_2})
\frac{e^{-Nx (\cosh s_1+ \cosh s_2) }
+e^{-Nx(\cosh s_2 + \cosh s_1)}}
{-2Nx(\cosh s_1 + \cosh s_2)},
\ee
and finally obtain
\be
\partial_x [ x Z_{-1}(x)] = 8(K_0^2(Nx) + K_1^2(Nx)).
\ee
The solution of this equation is given by
\be
Z_{-1}(x) =8C(K_0^2(Nx) - K_1^2(Nx)),
\ee
in complete agreement with (\ref{zxdirect}).

\end{document}